\documentclass[11pt,a4paper]{article}
\usepackage{jcappub}
\pdfoutput=1
\usepackage{ifthen}
\usepackage{epsfig}

\def\eps{{\cal E}}

\newcommand{\vmin}{\ensuremath{v_\mathrm{min}}}

\newcommand{\bx}{\ensuremath{\mathbf{x}}}  
\newcommand{\bv}{\ensuremath{\mathbf{v}}}  

\newcommand{\be}{\begin{equation}}
\newcommand{\ee}{\end{equation}}
\newcommand{\beq}{\begin{equation}}
\newcommand{\eeq}{\end{equation}}
\newcommand{\vect}[1]{\boldsymbol{\rm #1}}

\newcommand{\beqra}{\begin{eqnarray}}
\newcommand{\eeqra}{\end{eqnarray}}

\title{Anisotropic dark matter distribution functions and impact on WIMP direct detection}

\def\mpi{Max-Planck-Institut f{\"u}r Kernphysik,\\ 
Saupfercheckweg 1, 69117 Heidelberg, Germany}
\def\goe{Institut f{\"u}r Theoretische Physik,\\
Friedrich-Hund-Platz 1, 37077 G{\"o}ttingen, Germany}
\def\sthlm{Oskar Klein Centre for Cosmoparticle Physics, 
Department of Physics,\\ 
Stockholm University, SE-10691 Stockholm, Sweden}

\author[a]{Nassim Bozorgnia,}
\author[b]{Riccardo Catena}
\author[a,c]{and Thomas Schwetz}
\affiliation[a]{\mpi}
\affiliation[b]{\goe}
\affiliation[c]{\sthlm}
\emailAdd{bozorgnia@mpi-hd.mpg.de}
\emailAdd{riccardo.catena@theorie.physik.uni-goettingen.de}
\emailAdd{schwetz@mpi-hd.mpg.de}

\abstract{Dark matter N-body simulations suggest that the velocity distribution of dark matter is anisotropic. In this work we employ a mass model for the Milky Way whose parameters are determined from a fit to kinematical data. Then we adopt an ansatz for the dark matter phase space distribution which allows to construct self-consistent halo models which feature a degree of anisotropy as a function of the radius such as suggested by the simulations. The resulting velocity distributions are then used for an analysis of current data from dark matter direct detection experiments. We find that velocity distributions which are radially biased at large galactocentric distances (up to the virial radius) lead to an increased high velocity tail of the local dark matter distribution. This affects the interpretation of data from direct detection experiments, especially for dark matter masses around 10~GeV, since in this region the high velocity tail is sampled. We find that the allowed regions in the dark matter mass--cross section plane as indicated by possible hints for a dark matter signal reported by several experiments as well as conflicting exclusion limits from other experiments shift in a similar way when the halo model is varied. Hence, it is not possible to improve the consistency of the data by referring to anisotropic halo models of the type considered in this work.}


\begin{document}
\maketitle

\section{Introduction}

Dark matter direct detection experiments search for nuclear recoil
events induced by the scattering of Weakly Interacting Massive
Particles (WIMPs) providing the dark matter halo of the Milky Way with
nuclei in underground detectors. In order to predict the signal in
such experiments for given dark matter particle physics properties it
is necessary to specify also its local density and velocity
distribution.  Very little is known about the details of the local
dark matter phase space density and this lack of knowledge introduces
significant uncertainty in the interpretation of data from dark matter
direct detection experiments \cite{Ullio:2000bf, Green:2000jg,
  Belli:2002yt, Green:2002ht, Vergados:2007nc, Fairbairn:2008,
  MarchRussell:2008dy, Vogelsberger:2008qb, Kuhlen:2009vh,
  McCabe:2010zh, Green:2010gw, Lisanti:2010qx, Fairbairn:2012zs,
  Pato:2012fw}. To overcome these problems halo independent methods
have been developed and applied to data \cite{Fox:2010bu, Fox:2010bz,
  McCabe:2011sr, Frandsen:2011gi, HerreroGarcia:2011aa,
  Gondolo:2012rs, HerreroGarcia:2012fu, DelNobile:2013cta,
  Bozorgnia:2013hsa}.

In this work we follow a different approach and use information from
kinematical data on the Milky Way to constrain the properties of the
dark matter phase space distribution, based on reasonable assumptions
motivated by the results of N-body simulations. Following
Ref.~\cite{Catena:2009mf} a parameterization for the mass distribution
of the dark matter as well as baryonic components of the Milky Way is
adopted and its parameters are determined by a fit to kinematical
data. In addition to the ansatz of a spherically symmetric dark matter
distribution, in Ref.~\cite{Catena:2011kv} it was further assumed that
the dark matter velocity distribution is isotropic. In that case the
local velocity distribution function can be uniquely determined by
using the Eddington equation \cite{BT}, see also
\cite{Chaudhury:2010hj, Bhattacharjee:2012xm} for similar
approaches. In this work we keep the assumption of spherical symmetry
but we consider anisotropic velocity distributions. We allow for
different velocity dispersions in the radial and tangential
directions, conventionally described by the anisotropy parameter
$\beta$. We construct a set of self-consistent halo models with a
functional form of $\beta(r)$ motivated by the results from N-body
simulations, e.g.~\cite{Hansen:2004qs, Wojtak:2008mg, Ludlow:2011cs,
  Lemze:2011ud, Sparre:2012zk, Wojtak:2013eia}. Using inversion
procedures generalized from the Eddington equation we determine the
corresponding phase space density. Those halo models are then used for
an analysis of current data from direct detection experiments in order
to investigate the impact of the astrophysical uncertainties
(including anisotropy) on the interpretation of direct detection
data. In particular, we discuss the status of the controversal hints
for WIMPs with masses in the 10~GeV range from the DAMA~\cite{DAMA},
CoGeNT~\cite{CoGeNT:2011}, CRESST-II~\cite{CRESST}, and 
CDMS-Si~\cite{CDMS-Si:2013} experiments, versus the constraints from
XENON100~\cite{XENON100}, XENON10~\cite{XENON10},
KIMS~\cite{Kim:2012rza}, CDMS-Ge~\cite{CDMS:2009, CDMS-lt}, and LUX~\cite{LUX:2013} in the light
of our anisotropic self-consistent halo models.

The outline of the rest of the paper is as follows. In section~\ref{sec:DF} we discuss how to build a self-consistent dark matter phase space distribution function which leads to anisotropy parameter profiles $\beta(r)$ with a shape motivated from N-body simulations, departing from a given dark matter density profile and gravitational potential. In section~\ref{sec:MW} we describe our mass model for the Milky Way as well as the kinematical data we fit in order to constrain the parameters of our model. Section~\ref{sec:bayes} contains details of the Bayesian analysis of the galactic data, results for the Milky Way mass model parameters are given, and we describe how we extract the velocity distribution from the fit. In section~\ref{sec:DD} we introduce dark matter direct detection and describe the data we use from the various experiments for our analysis. In section~\ref{sec:results} our results are presented, and the implications of anisotropy as well as the variation of the parameters of the galactic model are discussed. Furthermore, we comment on the importance of including baryonic components in the galactic model. We conclude in section~\ref{sec:conclusions}. Details on the calculation of the anisotropy parameter $\beta$ are given in the appendix.

\section{Self-consistent anisotropic dark matter distribution functions}
\label{sec:DF}

The distribution function $f$ of a collisionless spherically symmetric system in a steady-state can be expressed as a function of two integrals of motion only \cite{BT}: the relative energy per unit of mass $\eps=\Phi_0 - \Phi - (1/2) v^2$, where $\Phi$ is the total gravitational potential acting on the system and $\Phi_0$ its value at the boundary, and the modulus of the total angular momentum $L=xv\sin\eta$, where $\eta$ is the angle between the position vector $\bx$ and the velocity $\bv$ of the constituents of the system ({\it e.g.} stars, dark matter particles, etc\dots). For these systems the collisionless Boltzmann equation can be drastically simplified, taking the very compact form
\begin{equation}
\frac{\partial f}{ \partial \eps} \frac{d\eps}{d t} + \frac{\partial f}{ \partial L} \frac{dL}{d t} = 0\,.
\label{eq:boltz}
\end{equation}
Clearly, if $f_{1}(\eps,L)$ and $f_{2}(\eps,L)$ are two independent solutions of Eq.~(\ref{eq:boltz}), any linear combination of these distribution functions will obey the same equation. Isotropic distribution functions can be further simplified and expressed as a function of the relative energy only, {\it i.e.} $f(\eps,L)\equiv f(\eps)$. A distribution function is self-consistent if it can be univocally related to the underlying mass profile of the system and the total gravitational potential generated by the system itself and eventually other components\footnote{This definition of self-consistent distribution function generalizes the one which would apply to a self-gravitating system, where the mass profile is sufficient to determine both the total gravitational potential and the distribution function.}. To construct a self-consistent distribution function one has to solve for $f$ the integral equation
\begin{equation}
\rho(\bx) = \int d^3v \,f(\eps,L)
\label{eq:master}
\end{equation}
which relates the relative potential $\Psi=\Phi_0-\Phi$ (entering the definition of $\eps$) and the mass profile $\rho$ to the distribution function $f$. 

The aim of this section is to introduce simple self-consistent {\it anisotropic} dark matter distribution functions which exhibit the same degree of anisotropy as observed in N-body simulations. The anisotropy of a distribution function is quantified by the anisotropy parameter which is defined as
\begin{equation}
\beta(r) = 1 - \frac{\sigma_t^2(r;f)}{2\sigma_r^2(r;f)} \,,
\label{eq:beta}
\end{equation} 
where $\sigma_t(r;f)$ and $\sigma_r(r;f)$ are respectively the
tangential and radial velocity dispersions obtained from the
distribution function $f$. (If not strictly necessary we will omit in
the following the dependence of $\beta(r)$ on $f$.) When
$\sigma^2_t=2\sigma^2_r$ the distribution function is isotropic and
$\beta(r)=0$. Radial anisotropy corresponds to the configuration
$\sigma_t\ll\sigma_r$ which implies $\beta(r)\approx 1$, while in the
opposite limit, namely $\sigma_t\gg\sigma_r$, $\beta(r)\rightarrow
-\infty$ and the distribution is said to be tangentially
anisotropic. In general $\beta(r)$ is a function of the galactocentric
distance $r$.  In N-body simulations typically it grows from
approximately zero in the center of the halo up to a value of about
0.2~--~0.4 for $r$ larger than the Sun's position and then it remains
constant or mildly decreases approaching the edge of the Galaxy, see
e.g.~\cite{Wojtak:2008mg, Ludlow:2011cs, Lemze:2011ud}.\footnote{In
  our work we keep the assumption of a spherically symmetric halo,
  such that $\beta$ is a function of $r$ only. The anisotropy in
  triaxial halos has been investigated for instance in
  Refs.~\cite{Sparre:2012zk, Wojtak:2013eia}.} Below we present
several possibilities to construct self-consistent distribution
functions providing such a behaviour for $\beta(r)$.

\subsection{Constant-$\beta$ plus  Osipkov-Merritt distribution functions}
\label{sec:OM}

A self-consistent distribution function with constant anisotropy
parameter can be constructed from the ansatz
\begin{equation}
f_{\gamma}(\eps,L)=G(\eps)L^{2\gamma},
\label{eq:const}
\end{equation} 
where $G(\eps)$ is a generic function of the relative energy and $\gamma$ a real constant. As shown in the appendix, this distribution function has by construction $\beta(r)=-\gamma$. Starting from Eq.~(\ref{eq:master}) and assuming a distribution function of the form~(\ref{eq:const}), one can express the function $G(\eps)$ in terms of $\Psi$ and $\rho$. One finds\footnote{The convergence of this integral requires $\gamma>-1$.} \cite{1991MNRAS.253..414C}
\begin{equation}
G(\eps) = \frac{\sin((n-1/2-\gamma)\pi)}{\pi \lambda(\gamma)\left(\gamma+\frac{1}{2}\right)!} \frac{d}{d\eps}\int_0^{\eps} \frac{d^n\rho_1(\Psi)}{d\Psi^n} \frac{d\Psi}{(\eps-\Psi)^{\gamma+3/2-n}},
\label{eq:constinv}
\end{equation}
where
\begin{equation}
\rho_1 \equiv \frac{\rho}{r^{2\gamma}}\,;\qquad\qquad \lambda(\gamma) = 2^{\gamma+3/2}\pi^{3/2}\frac{\Gamma(\gamma+1)}{\Gamma(\gamma+3/2)}
\end{equation}
and the integer $n$ is defined by $n=[\gamma+1/2] + 1$, with $[\gamma+1/2]$ the largest integer less than or equal to $\gamma+1/2$. Finally 
\begin{equation}
\left(\gamma+\frac{1}{2}\right)! \equiv \left\{ 
\begin{array}{ll}
(\gamma+\frac{1}{2})(\gamma-\frac{1}{2})\dots(\gamma+\frac{3}{2}-n) & \qquad \textrm{for} \,\,\gamma>-1/2 \\
1 & \qquad \textrm{for} \,\,-1<\gamma\le -1/2\,.
\end{array}
\right. 
\end{equation}
Though this distribution function allows to introduce some degree of anisotropy in the description of the local population of Milky Way dark matter particles, a constant value for $\beta(r)$ seems a too crude approximation, since N-body simulations generically predict an anisotropy parameter growing with $r$, at least up to a certain value of the galactocentric distance.

A popular example of anisotropic distribution function associated with a growing $\beta(r)$ is the Osipkov-Merritt distribution function \cite{Osipkov, Merritt}. It is constructed from the ansatz 
\begin{equation}
f(\eps,L)\equiv f_{\textrm{OM}}(Q)\,;\qquad\qquad Q \equiv \eps - \frac{L^2}{2 r_a^2}
\label{eq:OM}
\end{equation} 
where $r_a$ is a reference radius. For $r\ll r_a$,  $Q\rightarrow \eps$ and the Osipkov-Merritt distribution function is isotropic. At galactocentric distances larger than $r_a$ this distribution exhibits some degree of radial anisotropy approaching the regime $\beta(r)=1$ for sufficiently large values of $r$. By construction the  Osipkov-Merritt distribution function is associated with the anisotropy parameter (see appendix)
\begin{equation}
\beta(r) = \frac{r^2}{r^2+r^2_a}\,.
\label{eq:OMbeta}
\end{equation}
Starting from Eq.~(\ref{eq:master}) and assuming a distribution function of the form~(\ref{eq:OM}) one can express $f(Q)$ as a function of $\Psi$ and $\rho$. One finds \cite{1991MNRAS.253..414C}
\begin{equation}
f_{\textrm{OM}}(Q) = \frac{1}{\sqrt{8}\pi^2} \frac{d}{dQ}\int_0^{Q} \frac{d\rho_2(\Psi)}{d\Psi} \frac{d\Psi}{\sqrt{(Q-\Psi)}}
\label{eq:OMinv}
\end{equation}
where
\begin{equation}
\rho_2 = \left(1 + \frac{r^2}{r^2_a} \right) \rho\,.
\end{equation}
The major limitation of the Osipkov-Merritt distribution function is that it leads to an anisotropy parameter growing with a rate much larger than what is observed in the N-body simulations. 

In this paper we propose as a benchmark for galactic dark matter searches a simple anisotropic distribution function constructed as a linear combination of a distribution function associated with a constant $\beta(r)$ and a distribution function of the Osipkov-Merritt type:
\begin{equation}
f(\eps,L) = w f_{\textrm{OM}}(Q) + (1-w) f_{\gamma}(\eps,L),
\label{eq:linear}
\end{equation}
where $w$ is a real constant weighing the relative contribution of the two terms in the linear combination. The advantage of a distribution function of the type~(\ref{eq:linear}) is that it can faithfully reproduce the behavior of $\beta(r)$ observed in N-body simulations without requiring complicated inversion procedures to relate the distribution function to $\Psi$ and $\rho$. Indeed, the first term in Eq.~(\ref{eq:linear}) can be expressed as in Eq.~(\ref{eq:OMinv}) while the second term can be written in the integral form of Eq.~(\ref{eq:constinv}). Both terms can be easily evaluated by means of a straightforward numerical integration. The correct overall normalization for $f$ is guaranteed by the weights $w$ and $(1-w)$ introduced in the linear combination. By properly choosing the three free parameter ($w$, $r_a$, $\gamma$) entering the definition of the distribution function proposed here one is able to generate from Eq.~(\ref{eq:linear}) different functions $\beta(r)$, see Eq.~(\ref{eq:betalinear}) below. In the left panel of Fig.~\ref{fig:beta} we show three curves $\beta(r)$ obtained from three different choices of the parameters ($w$, $r_a$, $\gamma$). Comparing this figure to Fig.~3 of Ref.~\cite{Ludlow:2011cs}, one can appreciate the effectiveness of the benchmark distribution function proposed here in reproducing the results of the N-body simulations. The two distributions characterized respectively by $w=0.2$, $r_a=20$ kpc and $\gamma=-0.17$ (blue dashed curve), and $w=0.05$, $r_a=20$ kpc and $\gamma=-0.05$ (red dotted curve) bracket in fact the uncertainties in the predictions of the N-body simulations, while the model $w=0.15$, $r_a=20$ kpc and $\gamma=-0.10$ (black dot-dashed curve) provides a good approximation to the best fit found in Ref.~\cite{Ludlow:2011cs}. From Eq.~(\ref{eq:linear}) one obtains the following expression for the anisotropy parameter 
\begin{equation}
\beta(r)= \left[1 - w\frac{\sigma_t(r;f_{\textrm{OM}})}{ \sigma_r(r;f)} \right]_{\textrm{term 1}} - \left[(1-w)\frac{\sigma_t(r;f_\gamma)}{\sigma_r(r;f)} \right]_{\textrm{term 2}}
\label{eq:betalinear}
\end{equation}
which involves two terms, one with $f_{\textrm{OM}}$ in the numerator and the other proportional to an integral of $f_\gamma$. Explicit expressions for $\sigma_t(r;f_{\textrm{OM}})$ and $\sigma_t(r;f_\gamma)$ can be obtained using the expressions given in the appendix. In the right panel of Fig.~\ref{fig:beta} we show the relative contribution to $\beta$ of the first and second term in Eq.~(\ref{eq:betalinear}). For comparison we also show the Osipkov-Merrit anisotropy parameter for $r_a=20$ kpc. It is the cancellation between these two terms which produces an anisotropy parameter in agreement with the outcome of the N-body simulations.

\begin{figure}[t]
\begin{minipage}[t]{0.49\linewidth}
\centering
\includegraphics[width=\textwidth]{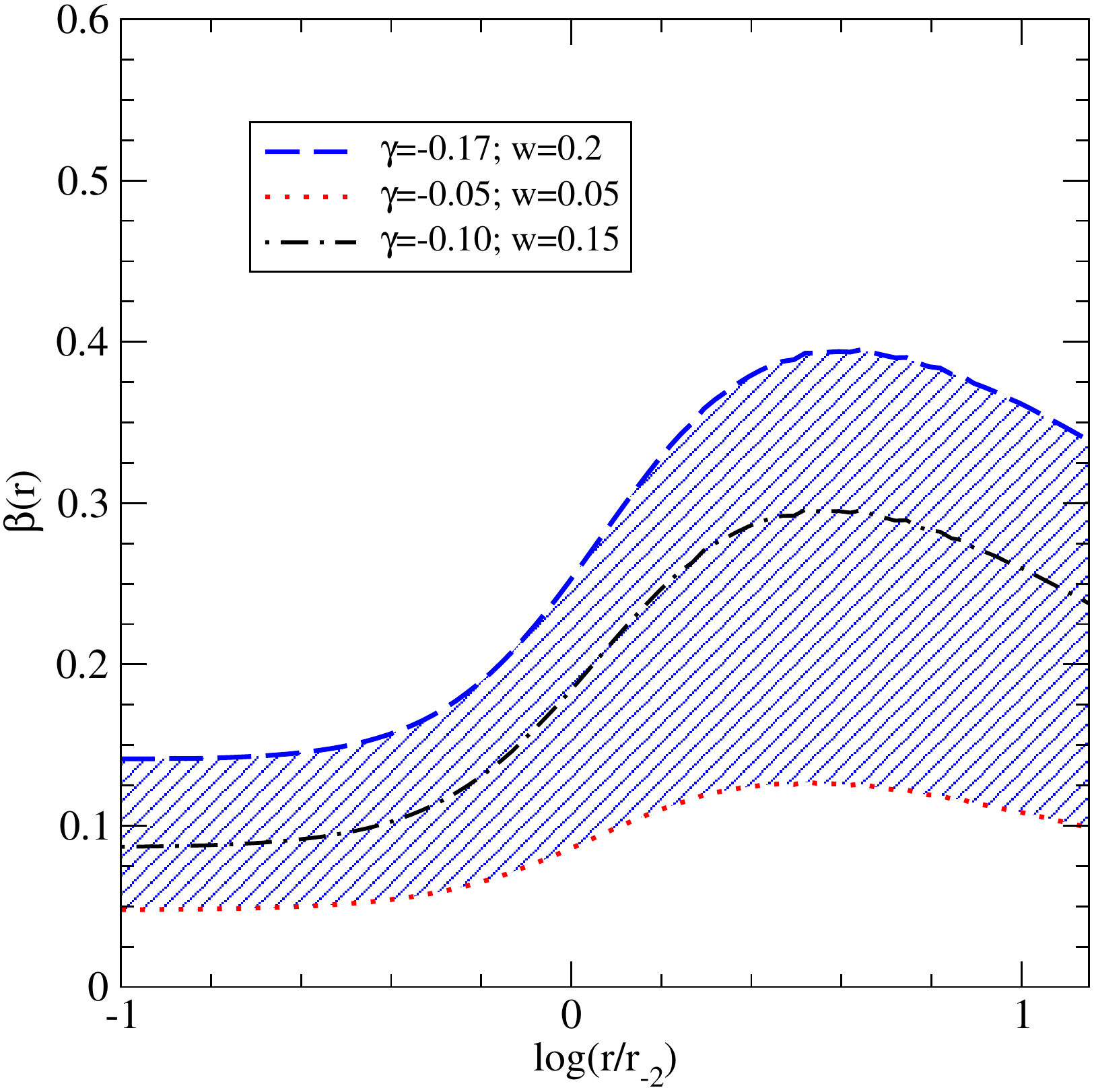}
\end{minipage}
\begin{minipage}[t]{0.49\linewidth}
\centering
\includegraphics[width=\textwidth]{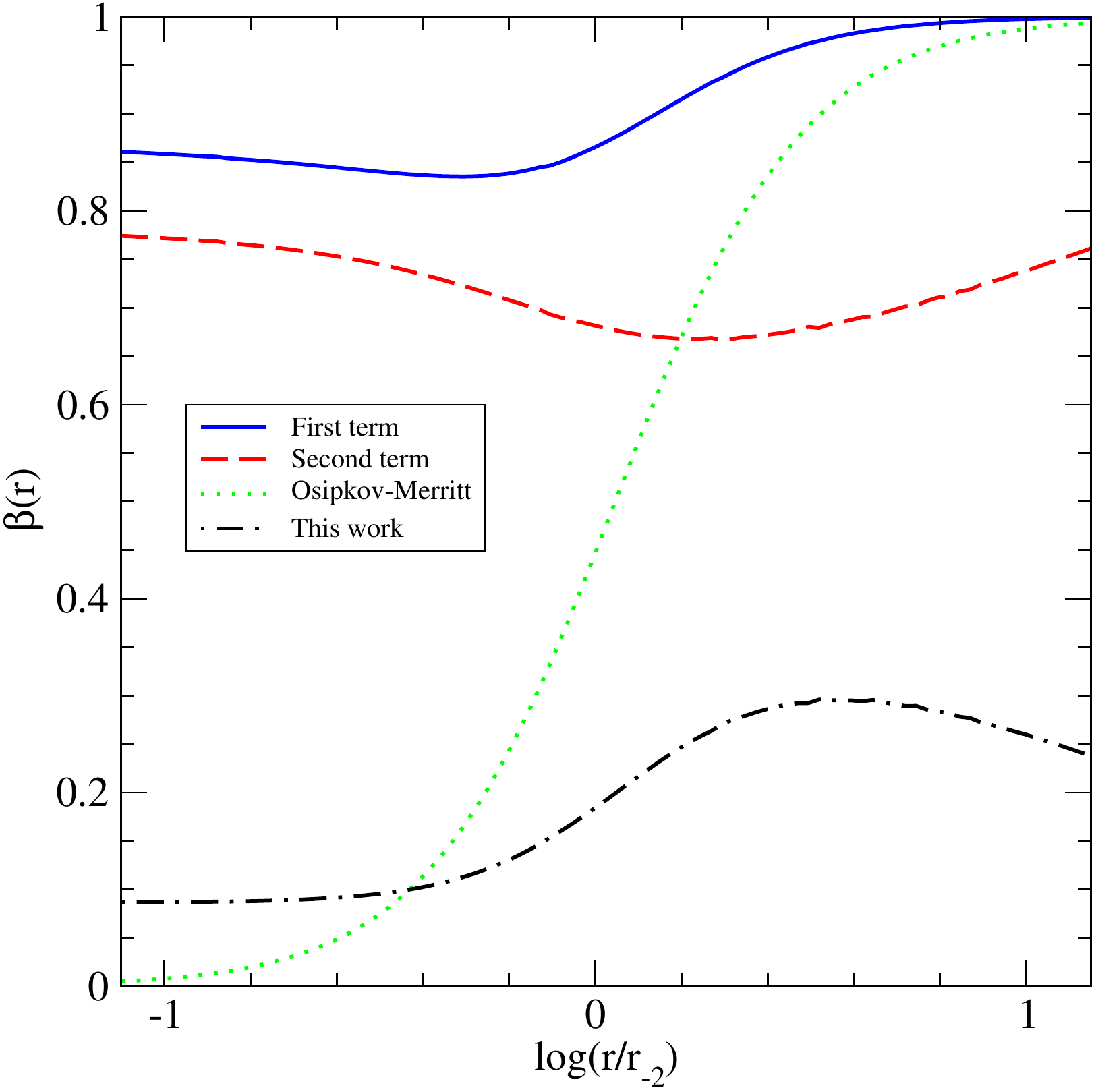}
\end{minipage}
\caption{Left panel: anisotropy parameter $\beta(r)$ as a function of the galactocentric distance $r$. Distinct curves correspond to different choices of the parameters ($w$, $r_a$, $\gamma$). The blue dashed curve corresponds to $w=0.2$, $r_a=20$ kpc, $\gamma=-0.17$, the red dotted curve to $w=0.05$, $r_a=20$ kpc, $\gamma=-0.05$, and the black dot-dashed curve to $w=0.15$, $r_a=20$ kpc, $\gamma=-0.10$. Right panel: Contributions to $\beta(r)$ of the first and second term in Eq.~(\ref{eq:betalinear}). For comparison we also show the Osipkov-Merrit case (with $r_a=20$ kpc) and the model proposed in this work and characterized by $w=0.15$, $r_a=20$ kpc and $\gamma=-0.10$.}
\label{fig:beta}
\end{figure}

\subsection{Alternative choices for constructing anisotropic halo models}
\label{sec:alt-models}

While the halo model of the previous subsection based on
Eq.~\eqref{eq:linear} allows to reproduce a $\beta(r)$ behaviour
similar to N-body simulations, the ansatz in Eq.~\eqref{eq:linear} is
certainly not unique. Therefore, we present in this subsection
alternative possibilities to construct halo models with similar
$\beta(r)$.

Let us write the dark matter distribution function as
\begin{equation}
  f(\eps, L) = k(\eps) h(\eps, L) \,.
\end{equation}
For $h(\eps, L)$ we make an explicit ansatz (see
below) in order to obtain anisotropy parameters $\beta(r)$ as
motivated by N-body simulations, while $k(\eps)$ is determined in order to fulfill the relation
\begin{equation}
  \rho(r) = \int d^3 v  f(\eps, L) 
\end{equation}
for the given $\rho(r)$ and $\Psi(r)$ obtained from the fit to the
Milky Way data described in the next section. We invert the 
function $\Psi(r)$ and consider $\Psi$ as the independent variable
instead of $r$. After changing the variables of integration one obtains
\begin{equation}\label{eq:voltera}
  \rho(\Psi) =  \int_0^\Psi d\eps \, k(\eps) \, K(\eps, \Psi)
\end{equation}
where 
\begin{align}
  K(\eps, \Psi) 
  &= 4 \pi \, \sqrt{2(\Psi - \eps)} \int_0^1 du \, h(\eps, L)
\end{align}
with $L = \sqrt{2(\Psi - \eps)(1-u^2)} r(\Psi)$. For a given
$h(\eps,L)$ the integral over $u$ is performed
numerically. Eq.~\eqref{eq:voltera} is a Voltera integral equation of
the second kind which we solve numerically in order to determine
$k(\eps)$.  For a given $h(\eps, L)$ and having obtained $k(\eps)$
from this procedure, we can calculate the anisotropy parameter by
numerically evaluating the integrals over the phase space density for
the calculations of $\sigma^2_t$ and $\sigma^2_r$ (see appendix for
explicit expressions).

\begin{table}
  \centering
  \begin{tabular}{ccccc}
\hline
    & $h(\eps, L)$ & $r_\kappa / r_{\rm vir}$ & $a$ & $b$ \\
\hline
case 1 & $h_B$ & 0.02  & 0.5 & 0.4 \\
case 2 & $h_A$ & 0.015 & 0.5 & 0.37 \\
case 3 & $h_A$ & 0.02  & 0.5 & 0.5 \\
case 4 & $h_B$ & 0.07  & 0.5 & 0.95 \\
\hline
  \end{tabular}
  \caption{Parameters for the 4 representative choices for $h(\eps,L)$. The expressions
   for $h_A$ and $h_B$ are given in Eqs.~\eqref{eq:hA} and \eqref{eq:hB}, 
   respectively. \label{tab:cases}}
\end{table}

\begin{figure}[t]
\centering
\includegraphics[width=0.5\textwidth]{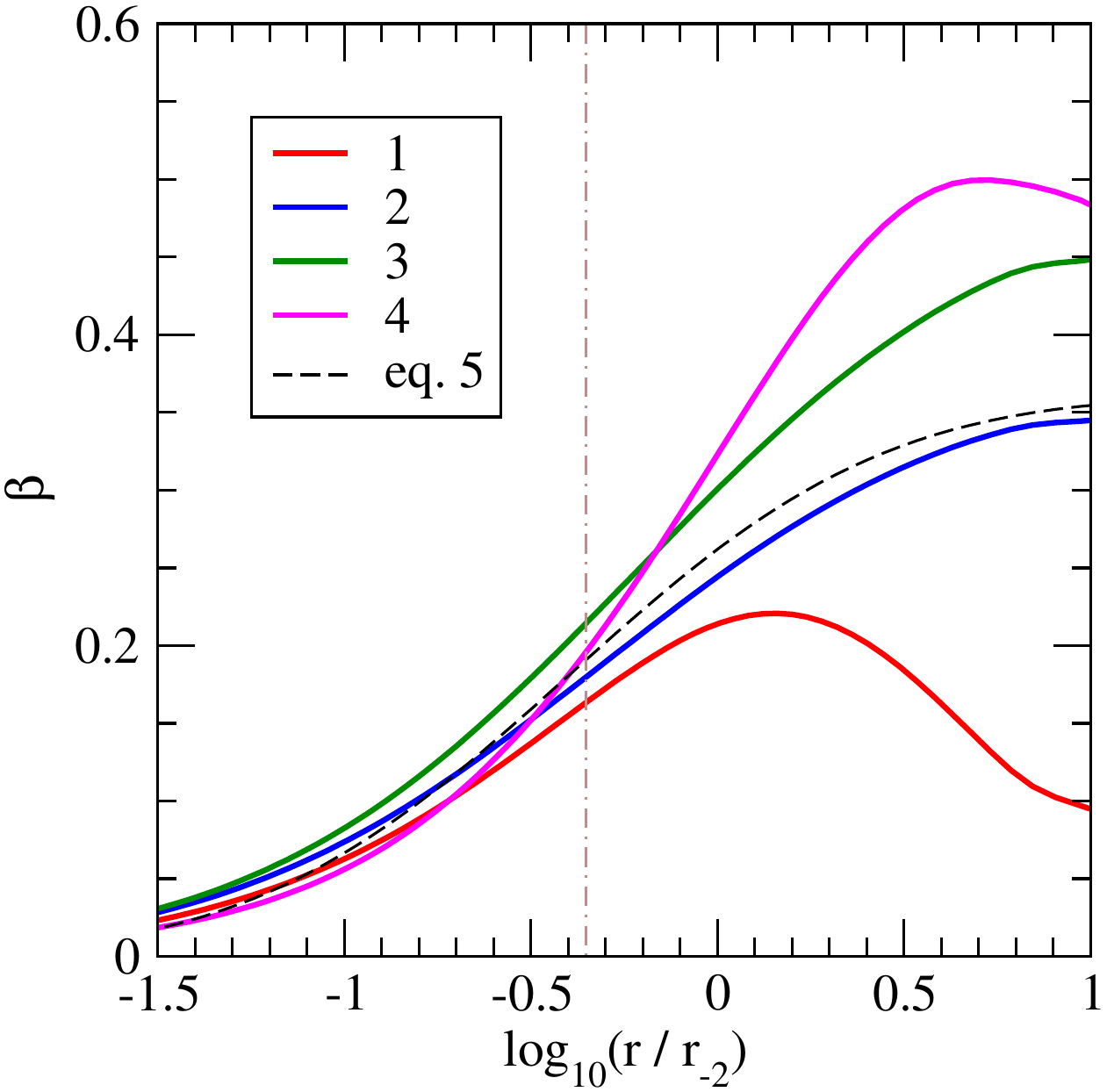}
\caption{Anisotropy parameter $\beta(r)$ as a function of the
  galactocentric distance $r$ for the four choices of $h(\eps, L)$ as
  specified in Tab.~\ref{tab:cases}. The dashed curve corresponds to
  Eq.~(5) of \cite{Ludlow:2011cs} (shown for comparison), and the vertical line indicates the location of the Sun at 8~kpc.
\label{fig:beta2}}
\end{figure}

Let us now specify $h(\eps, L)$. We consider here the two choices
\begin{align}
  h_A(\eps, L) &= (1 + \kappa)^{-b/a}  \,,\label{eq:hA} \\
  h_B(\eps, L) &= \left(1 + \kappa - \kappa e^{-10/\kappa}\right)^{-b/a} \,, \label{eq:hB}
\end{align}
where following \cite{1993MNRAS.261..283L, 1995MNRAS.275.1017C} we assume 
that $\eps$ and $L$ enter only through the particular dimensionless combination
\begin{align}
  \kappa = \left(\frac{L^2}{2 r_\kappa^2 \eps} \right)^a
\end{align}
with $r_\kappa$ being a constant characteristic radius. In Ref.~\cite{1993MNRAS.261..283L,
  1995MNRAS.275.1017C} the case $h_A$ with $a = 1$ is considered,
which is isotropic at small radii $ r \ll r_\kappa$ and assumes an
anisotropy parameter $\beta = b$ for $r \gg r_\kappa$.  Here we
consider 4 cases, with parameters as given in
Tab.~\ref{tab:cases}. Adopting a mass profile $\rho(r)$ and potential
$\Psi(r)$ from a fit to Milky Way data as described below we can
calculate for each ansatz the corresponding anisotropy parameter
$\beta(r)$ as shown in Fig.~\ref{fig:beta2}. The four choices for
$h(\eps, L)$ are motivated by the results of N-body simulations. The
four cases cover the spread of $\beta(r)$ shapes as reported for
instance in Ref.~\cite{Ludlow:2011cs} (see their Fig.~3, right panel).

\bigskip

To explicitly evaluate the distribution function and the associated
anisotropy parameter $\beta(r)$ one needs to specify a mass model for
the Milky Way from which to calculate the underlying mass profile
$\rho$ and the relative gravitational potential $\Psi$. The mass model
adopted in this work will be introduced in the next section. Let us
mention that when one tries to reconstruct the distribution function
from a given mass density and gravitational potental by
using inversion procedures as discussed above it is not guaranteed
that the resulting function $f(\eps, L)$ is non-negative everywhere,
as required as a physical consistency condition. We have checked that
the distribution functions we obtain always satisfy the condition
of being non-negative.

\section{A mass model for the Milky Way and kinematical data}
\label{sec:MW}

\subsection{Mass model for the Milky Way}
\label{sec:MW-model}

The mass model for the Milky Way adopted in the present analysis to evaluate the anisotropic dark matter distribution function introduced in the previous section has been extensively investigated in Refs.~\cite{Catena:2009mf,Catena:2011kv}. For completeness, we briefly summarize it in what follows, explicitly mentioning which parameters will be kept fixed and which will be instead considered as free parameters studying in the next section the impact of astrophysical uncertainties on the family of anisotropic distribution functions proposed here. 

The model consists of two luminous mass components, namely the stellar disk and the galactic bulge/bar component, and of a dark matter halo. Regarding the stellar disk, we assume a mass density profile which in cylindrical coordinates $(R,z)$ with origin in the galactic center is given by~\cite{Freudenreich:1997bx}
\beq
\rho_d(R,z) = \frac{\Sigma_{d}}{2 z_{d}} \, e^{-\frac{R}{R_d}} \, \textrm{sech}^2\left( \frac{z}{z_d}\right)
\;\;\;\; {\rm{with}} \;\;\;\; R<R_{dm}\;,
\label{disk}
\eeq 
where $\Sigma_{d}$ is the central disk surface density, $R_d$ and $z_d$ are length scales in the radial and vertical directions, while $R_{dm}$ is the truncation radius of the disk. $R_{dm}$ is assumed to scale with the local galactocentric distance $R_0$ according to the prescription $R_{dm}=12 \left[1+0.07(R_0-8~\textrm{kpc})\right]~{\rm kpc}$ and the vertical scale $z_{d}$ is fixed to the best fit value suggested in Ref.~\cite{Freudenreich:1997bx},  $z_{d}=0.340$~kpc. The bulge/bar region is instead characterized by the mass density profile~\cite{Zhao:1995qh}:
\beq
\rho_{bb}(x,y,z)= \bar{\rho}_{bb}\left[ s_a^{-1.85} \,\exp(-s_a) + \exp\left(-\frac{s_b^2}{2}\right) \right] \, ,
\label{bb}
\eeq
where 
\beq
s_a^2 =  \frac{q_b^2 (x^2+y^2)+z^2}{z_b^2},       
\eeq
and       
\beq
s_b^4 = \left[ \left(\frac{x}{x_b} \right)^2 + 
\left(\frac{y}{y_b} \right)^2\right]^2 +  \left(\frac{z}{z_b} \right)^4 \,.
\eeq
We implement in this analysis an axisymmetrized version of Eq.~(\ref{bb}), and assume $x_b \simeq y_b = 0.9~{\rm kpc} \cdot (8~{\rm kpc}/R_0)$, $z_b=0.4~{\rm kpc} \cdot (8~{\rm kpc}/R_0)$ and $q_b= 0.6$. See also \cite{Catena:2009mf} concerning the choice of these parameters. Rather than using the two mass normalization scales $\Sigma_{d}$ and $ \bar{\rho}_{bb}$ as free parameters, we re-parameterize these in terms of two dimensionless quantities, namely, the fraction of collapsed baryons $f_{\rm b}$ and the ratio between the bulge/bar and disk masses $\Gamma$:
\beqra
  f_{\rm b} &  \equiv & \frac{\Omega_{\rm DM}+\Omega_{\rm b}}{\Omega_{\rm b}} \frac{M_{bb}+M_d+M_{\textrm{H}_\textrm{I}}+M_{\textrm{H}_2}}{M_{vir}} \label{eq:barpar1} \\
  \Gamma & \equiv & \frac{M_{bb}}{M_d}\,.
  \label{eq:barpar2}
\eeqra
In Eq.~(\ref{eq:barpar1}) we also included the sub-leading contributions to the total virial mass $M_{vir}$ (defined in the following) associated with the atomic ($\textrm{H}_\textrm{I}$) and the molecular ($\textrm{H}_2$) galactic gas layers, with profiles as given in \cite{dame}. In summary, the free parameters describing the luminous components are $R_0$, $R_d$, $f_{b}$ and $\Gamma$.  

Concerning the dark matter halo component we consider an Einasto profile \cite{n04,graham}, which is favored by the latest N-body simulations and is given by 
\begin{equation}
  \rho(r)=\rho^{\prime} f_{E}\left(\frac{r}{r_{-2}}\right)\,,
  \label{nbody}
\end{equation}
with
\begin{equation}
  f_{E}(x) = \exp\left[-\frac{2}{\alpha} \left(x^{\alpha}-1\right)\right]\, , 
\label{eq:einasto}
\end{equation}
where $\alpha$ is a parameter controlling the slope of the profile. The reference normalization $\rho^{\prime}$ and the scale radius $r_{-2}$ in Eq.~(\ref{nbody}) are often rewritten as a function of the virial mass $M_{vir}$ and of the concentration parameter $c_{vir}$ by inverting the relations:
\beqra
  M_{vir} & \equiv & \frac{4\pi}{3} \Delta_{vir} \bar{\rho}_0\,R_{vir}^3 
  = \frac{\Omega_{\rm DM}+\Omega_{\rm b}}{\Omega_{\rm DM}} \,4\pi \int_0^{R_{vir}} dr \, r^2  \rho(r) \\
  c_{vir} & \equiv & R_{vir}/r_{-2},
\eeqra
where the virial overdensity $\Delta_{vir}$ in the first equation is computed according to Ref.~\cite{Bryan:1997dn} while $\bar{\rho}_0$ is the mean background density today. The presence in this equation of $\Omega_{\rm DM}$ and $\Omega_{\rm b}$, the dark matter and baryon energy densities in units of the critical density, reflects our assumption that only a fraction equal to ${\Omega_{\rm DM}}/({\Omega_{\rm DM}+\Omega_{\rm b}})$  of the total virial mass consists of dark matter. Their values have been set according to the mean values from the fit of the 7-year WMAP data~\cite{WMAP7} (employing here the latest Planck data \cite{Ade:2013zuv} would negligibly alter the present analysis). In the second equation, instead, $r_{-2}$ is the radius at which the effective logarithmic slope of the dark matter profile is equal to $-2$. Finally, we assume that the baryons which do not collapse in the disk are distributed according to the same profile as the dark matter component. The free parameters describing the dark matter halo are therefore $M_{vir}$, $c_{vir}$ and $\alpha$. The mass model used to compute the dark matter distribution function also includes an additional parameter, namely the anisotropy parameter $\beta_{\star}$ of a population of halo stars used in the analysis to constrain the model parameters. This additional parameter has been introduced in Ref.~\cite{Catena:2009mf} to include in the parameter estimation the velocity dispersion measurements of Ref.~\cite{SDSS6} (regarding $\beta_\star$ see also section~\ref{sec:dyncon}).

Given a mass model for the Milky Way one can calculate the associated gravitational potential solving the Poisson equation for $\Phi$ (or equivalently for the relative potential $\Psi$). A rigorous procedure would require the solution of partial differential equations in cylindrical coordinates, a method that would actually provide us with more information than those required in Eq.~(\ref{eq:linear}), which assumes a spherically symmetric $\Phi$. Moreover, the axisymmetric gravitational potential resulting from this procedure would be incompatible with the assumption $f \equiv f(\eps,L)$, which we made in section~\ref{sec:OM}, since $L$ is not in general an integral of motion of an axisymmetric system. We therefore have to approximate the true gravitational potential of our axisymmetric galactic model -- introduced to fit datasets which in many cases assume axial symmetry -- with a spherically symmetric $\Phi$, to proceed with our analysis consistently with the assumption $f \equiv f(\eps,L)$. We employ here the same approximation introduced in Refs.~\cite{Catena:2009mf,Catena:2011kv}, where the gravitational potential in the solar neighborhood is estimated as follows: first calculating the total mass profile $M(\bar{r})$ from the mass model defined in this section, {\it i.e.} the total mass $M(\bar{r})$ within a certain galactocentric distance $\bar{r}$; then using this quantity in the Poisson equation for $\Phi$ in the limit of spherical symmetry, whose solution can be written as
\begin{equation}
 \Phi(r) = \textrm{G}_{\rm N}\,\left[ \int_r^{R_{vir}}d\bar{r}\,\frac{M(\bar{r})}{\bar{r}^2} - \frac{M(R_{vir})}{R_{vir}} \right],
\label{eq:phi}
\end{equation}
where G$_{\rm N}$ is the Newton constant. This ``spherical symmetrization'' of $\Phi$ produces a gravitational potential compatible with the assumption $f \equiv f(\eps,L)$. We verified that the gravitational potential obtained by solving the Poisson equation for $\Phi$ in the limit of spherical symmetry (as explained above), and the {\it exact} gravitational potential of our model are sufficiently close to each other. To this aim we calculated the baryonic contribution to the axisymmetric potential of our model, $\Phi_{\rm bar}(R,z)$, using the appropriate integral solution of the Poisson equation, namely~\cite{KG89-571}
\begin{equation}
\Phi_{\rm bar}(R,z) = -2\pi G_{\rm N} \int_{0}^{+\infty} dk~J_{0}(kR)\int_{-\infty}^{+\infty} d\zeta~\tilde{\rho}_{\rm bar}(k,\zeta) \mathrm{e}^{-k|z-\zeta|},
\label{eq:axipot}
\end{equation}
where $J_0$ is a Bessel function and $\tilde{\rho}_{\rm bar}(k,z)$ is the Hankel transform of $\rho_{\rm bar}(R,z)\equiv\rho_{bb}(R,z)+\rho_d(R,z)$ in the R variable. 
\begin{figure}[t]
\begin{minipage}[t]{\linewidth}
\centering
\includegraphics[width=0.6\textwidth]{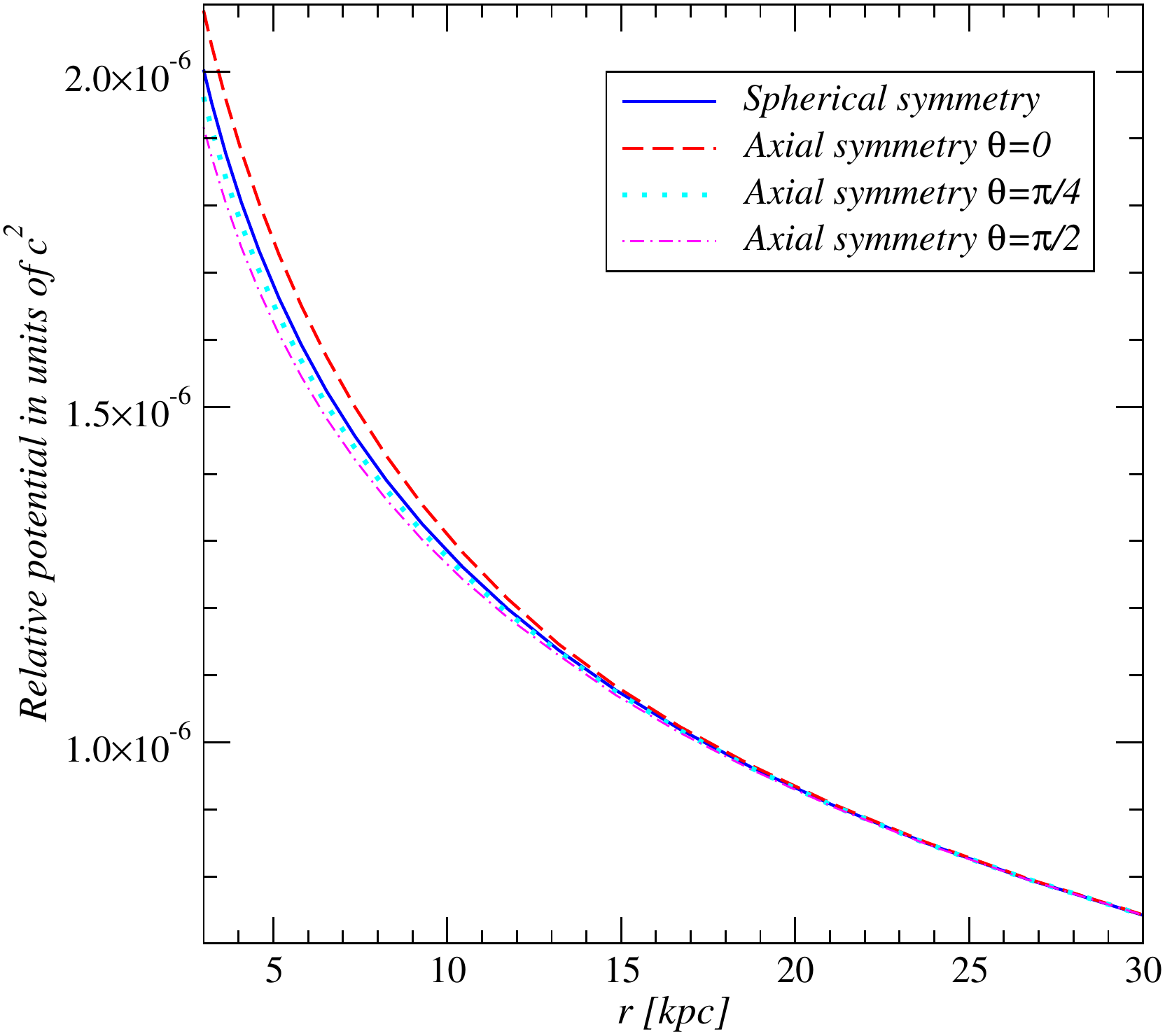}
\end{minipage}
\caption{Relative gravitational potential as a function of the galactocentric distance computed within the spherical approximation described in the text (blue line) and assuming axial symmetry according to Eq.~(\ref{eq:axipot}). For the axially symmetric case we show the relative potential profile along three directions forming an angle with the galactic plane of $\theta=0$  (red dashed line), $\theta=\pi/4$ (cyan dotted line) and of $\theta=\pi/2$  (magenta dot-dashed line) respectively.}
\label{fig:axipot}
\end{figure}
We find that in the solar neighborhood the two approaches provide results in agreement at the $\sim 3$ percent level, as one can see from Fig.~\ref{fig:axipot}, where the relative potential profile of our model is compared with the gravitational potential computed in the limit of spherical symmetry. Hence, we expect that our assumption of $L$ being an integral of motion should be fulfilled at the level of few percent. Alternative approximation schemes have been presented in Ref.~\cite{Widrow:2008yg}, where a Bayesian fit of the galaxy is performed assuming spherically symmetric halo and bulge and an axisymmetric disk, and in a second Bayesian study~\cite{Deg:2012eu} where a mass model for the Milky Way is fit to observations of the Sagittarius stream together with other photometric and kinematic data.

With this last step, one has all the ingredients to evaluate the anisotropic dark matter
distribution function~(\ref{eq:linear}). In summary, besides the three
parameters controlling the degree of anisotropy of $f$, namely $w$,
$\gamma$ and $r_a$, our benchmark distribution function depends on the
following 8 galactic model parameters: $R_0$, $R_d$, $f_{b}$, $\Gamma$
for the luminous components, $M_{vir}$, $c_{vir}$, $\alpha$ for the
dark matter halo, and $\beta_{\star}$ for halo stars.

\subsection{Dynamical constraints}
\label{sec:dyncon}

The distribution function~(\ref{eq:linear}) depends on 11 parameters, three fixed requiring a good agreement with the N-body simulations, namely $w$, $\gamma$ and $r_a$, and 8 galactic model parameters subject to a variety of complementary constraints derived from different observations of the Milky Way properties. These constraints will be used in the next section to determine the uncertainties within which the galactic model parameters are known. This information will allow us to determine the impact of astrophysical uncertainties on the benchmark distribution function studied in this paper.

There are different classes of constraints which are relevant for the present study and will be therefore implemented in the Bayesian analysis described in the next section. A first class of constraints concerns the direct measurement of the kinematical properties of different tracers of the Milky Way gravitational potential. Terminal velocities, namely the extreme velocities observed monitoring the motion of $\textrm{H}_\textrm{I}$ and CO gas clouds along different line of sights, have been often used in the literature to constrain the Milky Way rotation curve at galactocentric distances smaller than the Sun's position. Here we adopt the compilation of terminal velocities published in Ref.~\cite{mal} consisting of 111 terminal velocities. We then compare these observations with the theoretical prediction of our mass model for the Milky Way, namely
\begin{equation}
v_{t}(r) =  v_c(r) - v_c(R_0) \frac{r}{R_0},
\label{eq:terminal}
\end{equation} 
where $v_c(r)$ is the circular velocity, {\it i.e.} the rotation curve, which, assuming spherical symmetry, is given by $v^2_c(r)=r d\Phi/dr$. Another population of tracers which has recently played a major role in the context of galactic mass modeling is the population of about 2400 Blue-Horizontal-Branch halo stars selected from the SDSS DR-6 for which accurate kinematical properties have been published in Ref.~\cite{SDSS6}. Here we compare the observed radial velocity dispersion $\sigma_{r}(r)$ of this tracer population (we employ the 9 data points of Fig.~10 in Ref.~\cite{SDSS6}) with the expectations of our galactic model, which under the assumption of spherical symmetry both for the mass profile of the tracer population and for the total gravitational potential predicts 
\begin{equation}
\sigma_r^2(r) = \frac{1}{r^{2\beta_\star}\,\rho_\star(r)} \int_r^{\infty}d\tilde{r}\;\tilde{r}^{2\beta\star} \rho_\star(\tilde{r}) \frac{d\Phi}{d\tilde{r}}, 
\label{eq:BHB}
\end{equation} 
where $\rho_\star \propto r^{-3.5}$ is the halo star density and $\beta_\star$ the constant anisotropy parameter of this stellar system, treated as explained above, as a free parameter in the analysis performed in the next section. 

A second class of constraints regards the observation of ``integrated properties'' of the Milky Way obtained integrating along the line of sight, or over certain portions of the three-dimensional space, the mass profiles of the different galactic components. The total mass of the Milky Way within 50 kpc and 150 kpc, measured observing the motion of the Milky Way satellites or the radial velocity of distant halo stars and obtained from our galactic model integrating the total mass density within the corresponding volumes, will be used in this work to constrain the parameters affecting the total mass of the Milky Way. We will make use here of the results 
\begin{equation}
M(<50~\textrm{kpc}) = (5.4\pm0.25)\times10^{11} M_\odot
\label{eq:50}
\end{equation} 
from Ref.~\cite{Sakamoto} and
\begin{equation}
M(<150~\textrm{kpc}) =  (7.5\pm2.5)\times10^{11} M_\odot
\label{eq:150}
\end{equation} 
from Ref.~\cite{Deason:2012ky}. The latter constraint is in agreement with recent measurements of the total mass of the Milky Way which find a value for the mass of our Galaxy approximately a factor of 2 lower than previously expected (see Ref.~\cite{Deason:2012ky} and references therein). Another powerful constraint belonging to this class is the total mean surface density within 1.1 kpc, $\Sigma_{|z|<1.1 \textrm{kpc}}$, which has been reexamined in various analyses in recent years (see for instance Ref.~\cite{Bovy:2013raa}) always producing results in agreement with the original work of Kuijken \& Gilmore who studying the vertical motion of a population of $K$ halo stars found~\cite{KG91}
\begin{equation}
\Sigma_{|z|<1.1 \textrm{kpc}} = (71\pm6) \,M_\odot \,\textrm{pc}^{-2} \,.
\label{eq:sigma11}
\end{equation} 
We will implement this value in our analysis together with the constraint on the local surface density corresponding to the visible components, $\Sigma_\star$, which has been instead estimated with star counts~\cite{KG89}
\begin{equation}
\Sigma_\star = (48\pm8) \,M_\odot\, \textrm{pc}^{-2} \,.
\label{eq:sigmastar}
\end{equation} 
In this context, the observation of microlensing events along certain specific directions pointing towards the bulge region has the capability of imposing interesting constraints on the normalization of the mass profile of the luminous galactic components \cite{Iocco:2011jz}. The probability of observing one of these events is related to the so-called optical depth of the region of interest. Given a mass model for the Milky Way, this can be calculated as follows~\cite{Iocco:2011jz}
\begin{equation}
\tau(\ell,b)= \frac{4\pi G}{N c^{2}}\int_0^{r_\infty} dD_s \frac{dn_s}{dD_s} \int_{0}^{D_s} dD_l \rho_l(\ell,b,D_l) D_l\left(1-\frac{D_l}{D_s}\right)
\label{eq:tau1}
\end{equation}
where $D_s$ is the distance between the observer and the sources involved in these microlensing events, namely the material forming the bulge/bar region, while $D_l$ is the distance between the observer and the corresponding ``lenses'', which in this study are made of the material forming both the bulge/bar region and the stellar disk, whose density is given by $\rho_l = \rho_{bb}+\rho_d$. $dn_s/dD_s\propto \rho_s(D_s) D_s^{2}$ is the distance distribution of the detectable sources, $\rho_s = \rho_{bb}$ and $N = \int_0^{r_\infty} dD_s \,dn_s/dD_s$ with $r_{\infty} = 20$ kpc. We adopt here the 2005 measurement of $\tau$ made by the MACHO collaboration~\cite{Popowski:2004uv}:
\begin{equation}
\tau(\bar{\ell},\bar{b})=2.17^{+0.47}_{-0.38} \times 10^{-6}  \quad \textrm{with}  \quad (\bar{\ell},\bar{b}) = (1.50^{\circ}, -2.68^{\circ}) \,.
\label{eq:tau2}
\end{equation}

Finally, we have employed in the present analysis constraints obtained from the measurement of local properties of the Milky Way rotation curve, conveniently encoded in two linear combinations of Oort's constants, namely $A+B$ and $A-B$, and the value of the local circular velocity $v_c(R_0)$. The sum of the Oort's constants A and B is proportional to the local slope of the galactic rotation curve, {\it i.e.}
\begin{equation}
A+B= -\left(\frac{\partial v_c}{\partial R}\right)_{R=R_0}\,,
\label{eq:ApB1}
\end{equation}
while the difference of these constants gives 
\begin{equation}
A-B= \frac{v_c(R_0)}{R_0} \,.
\label{eq:AmB1}
\end{equation}
There are still great uncertainties in the combination of Oort's constants $A+B$. We will implement in our analysis a value derived from Ref.~\cite{Fuchs:2009mk}. This has been found studying the kinematics of a population of old M type stars of the thin disk selected from the SDSS data and it is compatible with zero within one standard deviation:
\begin{equation}
A+B=(0.18\pm0.47)\,\textrm{km}\,\textrm{s}^{-1}\,\textrm{kpc}^{-1} \,.
\label{eq:ApB2}
\end{equation}
Different techniques have been instead used in the literature to estimate $A-B$. These range from the study of the motion of various populations of stars in the solar neighborhood to the observation of the apparent motion of the radio source Sgr A$^{*}$, which is believed to trace the position of the massive black hole at the center of the Milky Way. In this context it has been recently claimed~\cite{Bovy:2012ba} that this latter measurement, which found \cite{reid2} $A-B=(29.45\pm0.15)\,\textrm{km}\,\textrm{s}^{-1}\,\textrm{kpc}^{-1}$, should be corrected taking into account that the offset between the so-called local standard of rest (LSR), namely the velocity of a circular orbit passing at the Sun's position, and the actual local rotational velocity of the Sun is larger than previously expected. This correction led to an estimate of $A-B$ in perfect agreement with the accurate determination of the same quantity made by the Hipparcos satellite, which will be therefore implemented in the next section as a constraint on $A-B$. This determination has a significantly lower central value and reads as follows~\cite{1997MNRAS.291..683F}
\begin{equation}
A-B=(27.2\pm0.9)\,\textrm{km}\,\textrm{s}^{-1}\,\textrm{kpc}^{-1}\,.
\label{eq:AmB2}
\end{equation}
We want to stress here that the combination of Oort's constant $A-B$ is very important in the determination of local quantities relevant for dark matter direct detection. It is for instance positively correlated with the local dark matter density  $\rho_{\text{loc}}$ (see Fig.~\ref{fig:derivs}), {\it i.e.} the larger is $A-B$ the higher is  $\rho_{\text{loc}}$. Indeed, the mean value found here for the local dark matter density having assumed the constraint (\ref{eq:AmB2}) (as well as the low value of $v_c(R_0)$ in Eq.~(\ref{eq:vc}), see below) is close to 0.3$\,\textrm{GeV}\,\textrm{cm}^{-3}$, in agreement with independent analyses of $\rho_{\text{loc}}$ based on similar datasets~\cite{Bovy:2012tw}. This value is lower than what we would have found assuming instead $A-B=(29.45\pm0.15)\,\textrm{km}\,\textrm{s}^{-1}\,\textrm{kpc}^{-1}$ (and $v_c(R_0)\sim245\,\textrm{km}\,\textrm{s}^{-1}$~\cite{Catena:2009mf}). Indeed, as independently shown by various groups the latter value for $A-B$ (together with a larger $v_c(R_0)$) would have led to $\rho_\textrm{loc}\sim 0.4\,\textrm{GeV}\,\textrm{cm}^{-3}$. Though the focus of this work is on the impact of anisotropic distribution functions on the direct detection of WIMPs and not on the determination of the local dark matter density (see for instance Refs.~\cite{Catena:2009mf,Salucci:2010qr,Pato:2010yq,deBoer:2010eh,Iocco:2011jz,Garbari:2012ff,Bovy:2012tw} for a discussion on this subject), we underline here that it will be crucial in the future to accurately determine $A-B$, convincingly establishing whether or not the offset between the LSR and the actual local Sun's rotational velocity is as large as quoted in Ref.~\cite{Bovy:2012ba}. 

Having assumed in the present analysis the constraint~(\ref{eq:AmB2}), we will consistently implement in our study the determination of the local circular velocity of the Sun found in Ref.~\cite{Bovy:2012ba}, namely 
\begin{equation}
v_c(R_0)=(218\pm6)\,\textrm{km}\,\textrm{s}^{-1}\,.
\label{eq:vc}
\end{equation}
This value of $v_c(R_0)$ was found analyzing 3365 stars selected from the first year of data of the Apache Point Observatory Galactic Evolution Experiment (APOGEE) and it is lower than what was found in a previous work~\cite{Reid:2009nj} studying the motion of masers associated with star forming regions located at galactocentric distances larger than the Sun's position. Again following Ref.~\cite{Bovy:2012ba}, we do not include in the present analysis the estimate of $v_c(R_0)$ made using these masers because a refined study of this tracer population \cite{Bovy:2009dr} found that these objects can significantly constrain the rotation curve only when a large number of prior assumptions are made in the data analysis. Moreover, the higher value of $v_c(R_0)$ found with this approach might reflect a bias in the data related to the fact that this tracer population is lagging with respect to circular motion by about 15 km s$^{-1}$, an offset which, according to Ref.~\cite{Bovy:2009dr}, is quite large for a young and relatively cold tracer population. As for the combination of Oort's constants $A-B$, it will be very important to clarify in the near future which is the correct value of the local circular velocity. 

\section{Bayesian analysis}
\label{sec:bayes}

Having introduced our ansatz for the anisotropic velocity distribution function, the galactic model on which it depends, and the constraints acting on the associated 8-dimensional parameter space, we can now focus on the Bayesian analysis of this distribution function. The aim is to determine the regions in the 8-dimensional parameter space favored by the data and then to extract from this information the anisotropic distribution function observationally favored within this setup and the corresponding uncertainties. In a Bayesian framework this corresponds to determining the posterior probability density function (PDF) characterizing our 8-dimensional parameter space. This PDF will then allow us to determine the PDF associated with generic functions of the galactic model parameters, including the benchmark distribution function proposed in this work. 

According to Bayes' theorem the posterior PDF of certain parameters -- conveniently grouped in an array ${\bf p}$ -- is proportional to the product of the Likelihood function $\mathcal{L}(\bf{p,d})$ and the prior probability density $\pi({\bf p})$:
\beq
\mathcal{P}({\bf p,d}) = \frac{\mathcal{L}({\bf d,p}) \pi({\bf p})}{ E({\bf d})}
\label{eq:bayes}
\eeq
where ${\bf d}$ is the array of datasets used to constrain the parameter space. The Bayesian evidence $E({\bf d})$, being independent from ${\bf p}$, plays the role of a normalization constant when performing parameter inference. The marginal posterior PDF of a generic function $g$ of the parameters ${\bf p}$, {\it e.g.}\  in our case the dark matter distribution function (evaluated at a certain velocity), is given by the expression
\beq
p(g | {\bf d}) = \int d{\bf p}\,\delta(g({\bf p})-g)\,\mathcal{P}({\bf p}|{\bf d}) \,,
\label{eq:gpdf}
\eeq
which follows from the definition of conditional probability. 

\begin{figure}[t]
\centering
\includegraphics[width=\textwidth]{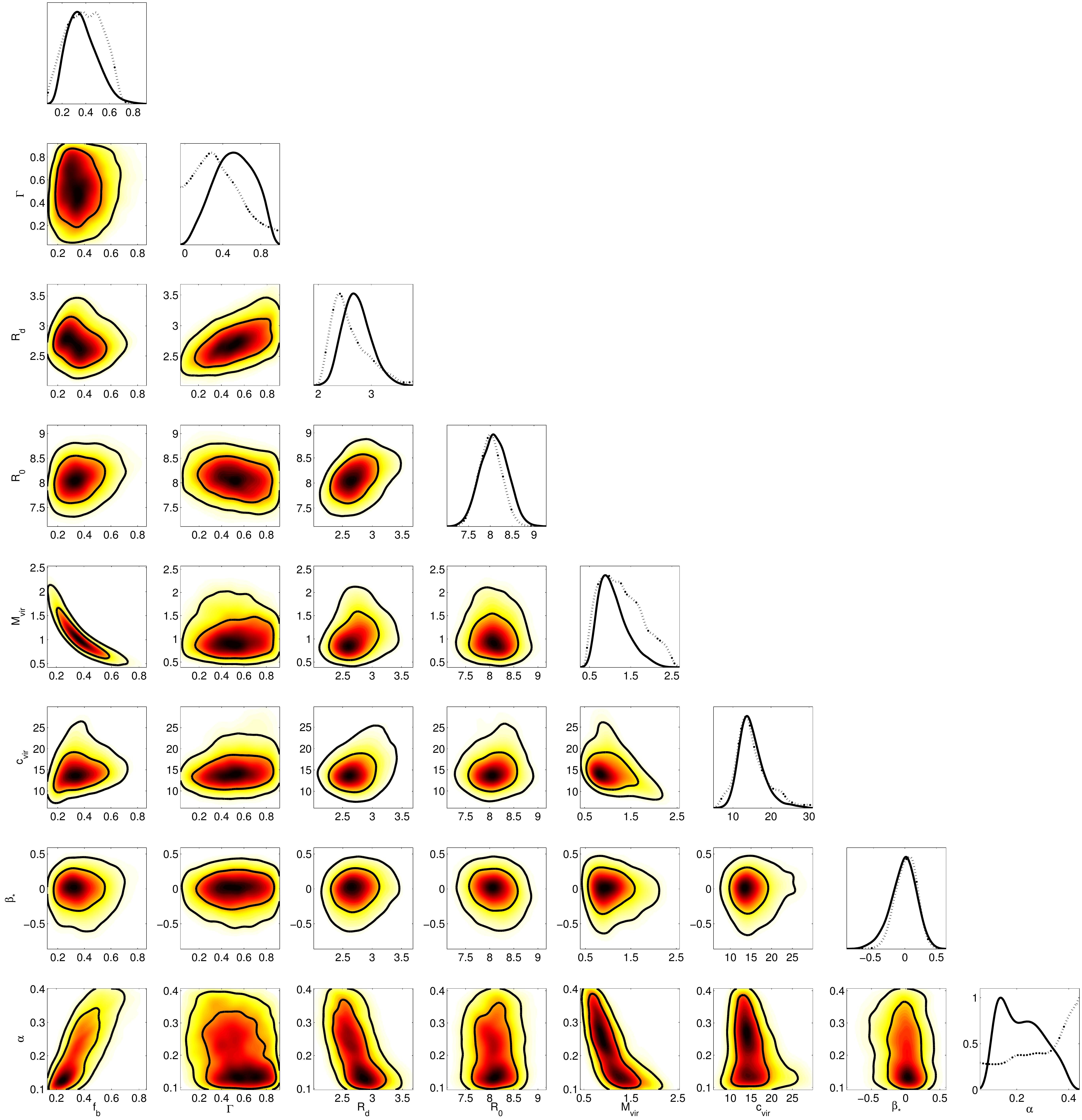}
\caption{1- and 2-dimensional marginal posterior PDFs for the galactic model parameters. In the off-diagonal panels the black curves enclose the 68\% and 95\% credible regions while in the diagonal panels the black curve is the 1-dimensional PDF and the dotted line corresponds to the mean Likelihood (see appendix C of Ref.~\cite{Lewis:2002ah} for a definition).}
\label{fig:params}
\end{figure}

The form of the likelihood function implemented in the present analysis is a multivariate Gaussian distribution. Each observable contributes to the Likelihood through a Gaussian factor characterized by the means and standard deviations reported in section~\ref{sec:dyncon}. Concerning the choice of the prior probability density $\pi({\bf p})$, we consider flat priors for all the parameters of our galactic model. A test of the dependence of the results from the priors in the context of a similar analysis has been performed in Ref.~\cite{Catena:2009mf}, where it is shown that when constraining the underlying mass model the Likelihood is more informative than the assumed prior PDF.

\begin{figure}[t]
\centering
\includegraphics[width=\textwidth]{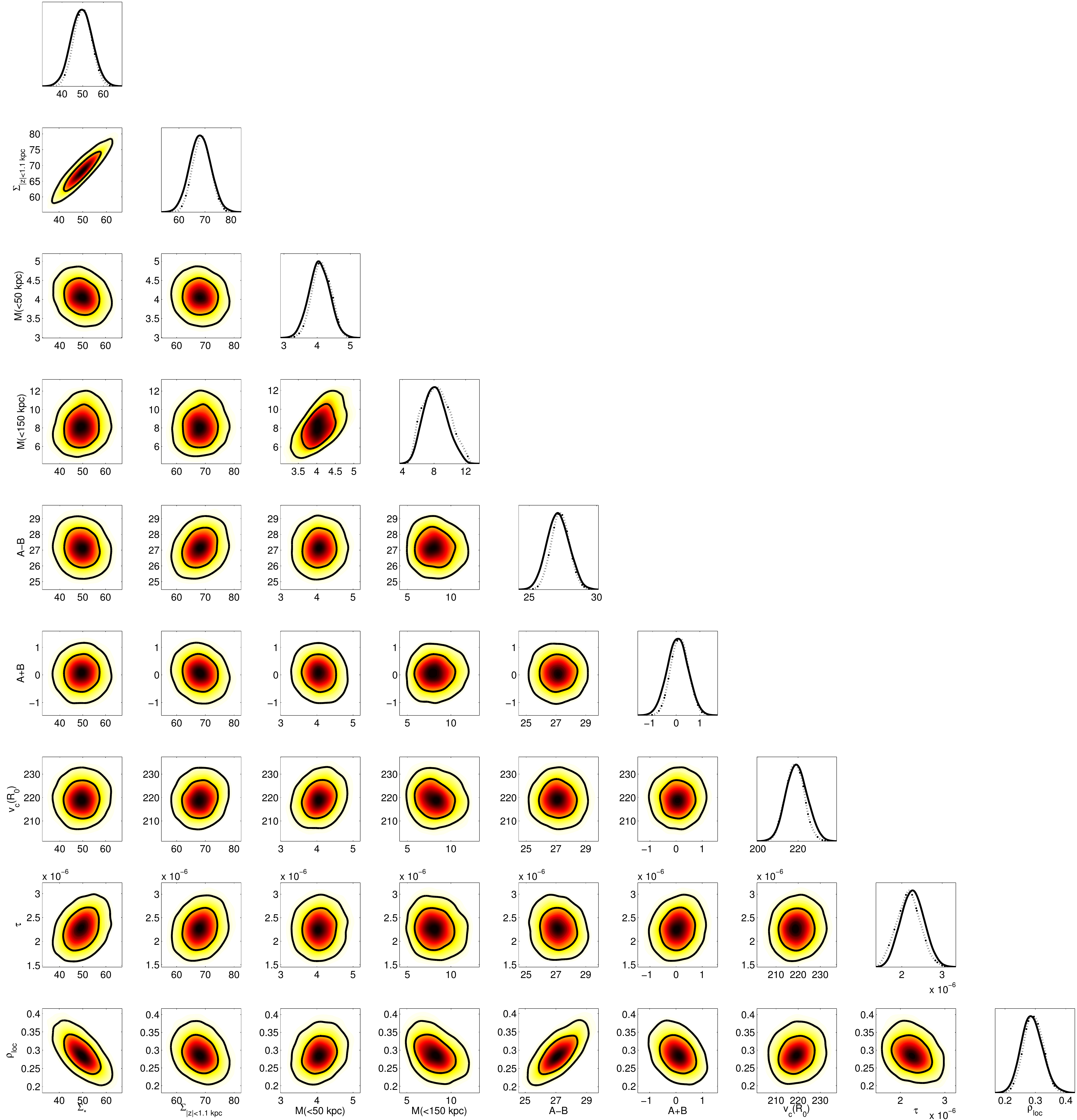}
\caption{1- and 2-dimensional marginal posterior PDFs for selected functions of the galactic model parameters. The notation is the same as in Fig.~\ref{fig:params}. }
\label{fig:derivs}
\end{figure}

Regarding the galactic model parameters, we present results in terms of 1- and 2-dimensional marginal posterior PDFs, which are constructed integrating the full posterior PDF over the six or seven remaining dimensions. In Fig.~\ref{fig:params} we show these PDFs for our galactic model parameters. All parameters can be reconstructed within the setup considered here, including the parameters associated with the bulge/bar region whose mass profile is constrained by the observation of microlensing events performed by the MACHO collaboration. We also present 2-dimensional marginal posterior PDFs for pairs of functions of the model parameters obtained analogously to the 1-dimensional PDF in Eq.~(\ref{eq:gpdf}). These are shown in Fig.~\ref{fig:derivs} to emphasize various correlations relating these quantities. As already mentioned it is particularly relevant in the context of dark matter searches the positive correlation observed between $A-B$ and $\rho_\textrm{loc}$, the local dark matter density, as well as the known correlation between $\rho_\textrm{loc}$ and $\Sigma_\star$. Adopting the constraint on $A-B$ reported in Eq.~(\ref{eq:AmB2}) and the estimate of $v_c(R_0)$ in Eq.~(\ref{eq:vc}), we find the following mean value for the local dark matter density: $\rho_{loc}=0.29\pm0.035\,\textrm{GeV}\,\textrm{cm}^{-3}$, where the reported error corresponds to the standard deviation. The 68\% (95\%) credible interval associated with $\rho_{loc}$ is $[0.25,0.32]\,\textrm{GeV}\,\textrm{cm}^{-3}$ ($[0.22,0.36]\,\textrm{GeV}\,\textrm{cm}^{-3}$).  

Let us emphasize that the results presented in Figs.~\ref{fig:params} and \ref{fig:derivs} (including the value of $\rho_{\rm loc}$) are independent of any assumption on the dark matter velocity distribution, in particular on the anisotropy. Indeed, the observables at disposal provide information only on the spatial distribution of the dark matter component and not on its distribution in velocity space. This requires additional assumptions as discussed in detail in section~\ref{sec:DF}.

\begin{figure}[t]
\begin{minipage}[t]{0.49\linewidth}
\centering
\includegraphics[width=\textwidth]{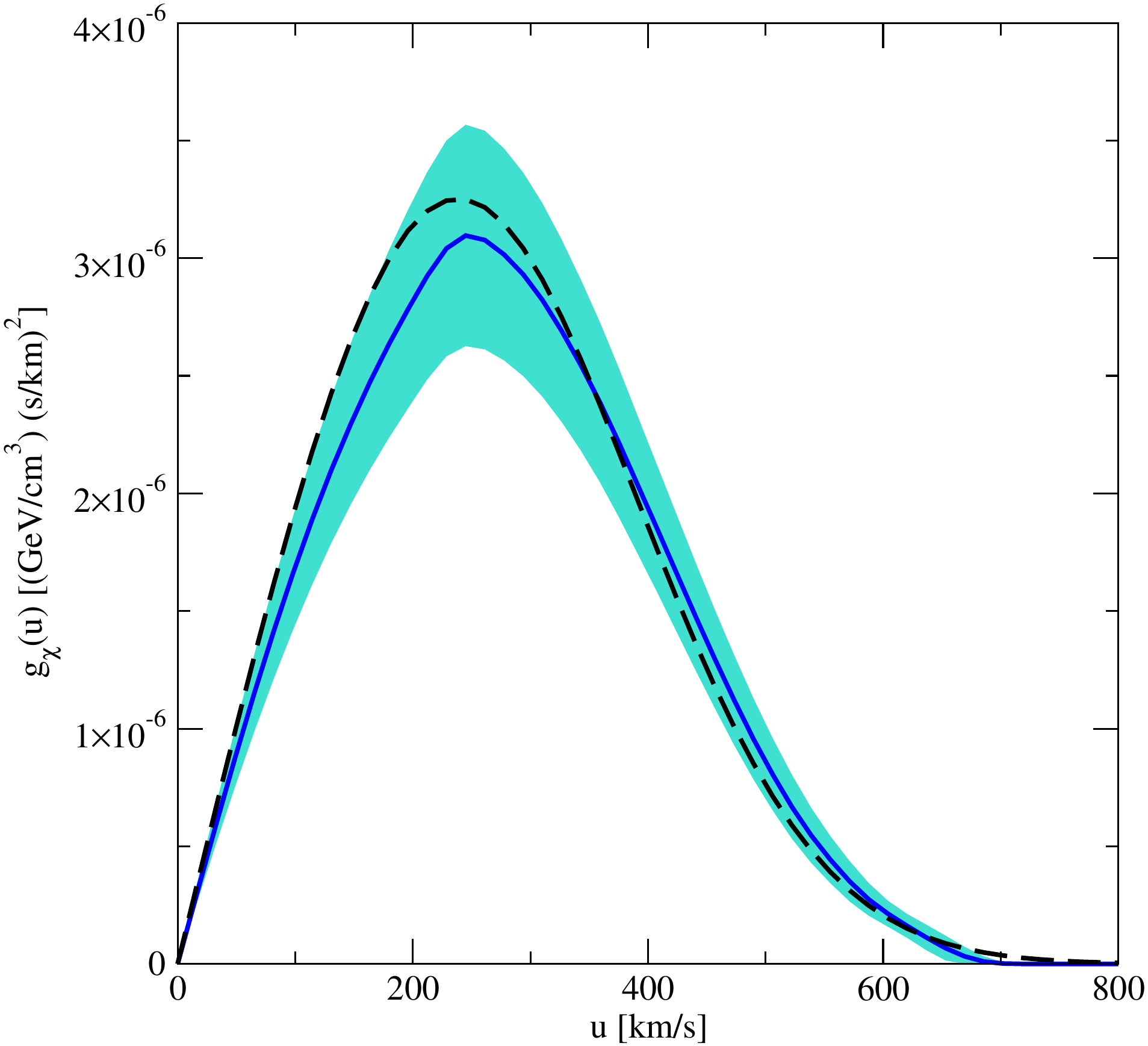}
\end{minipage}
\begin{minipage}[t]{0.49\linewidth}
\centering
\includegraphics[width=\textwidth]{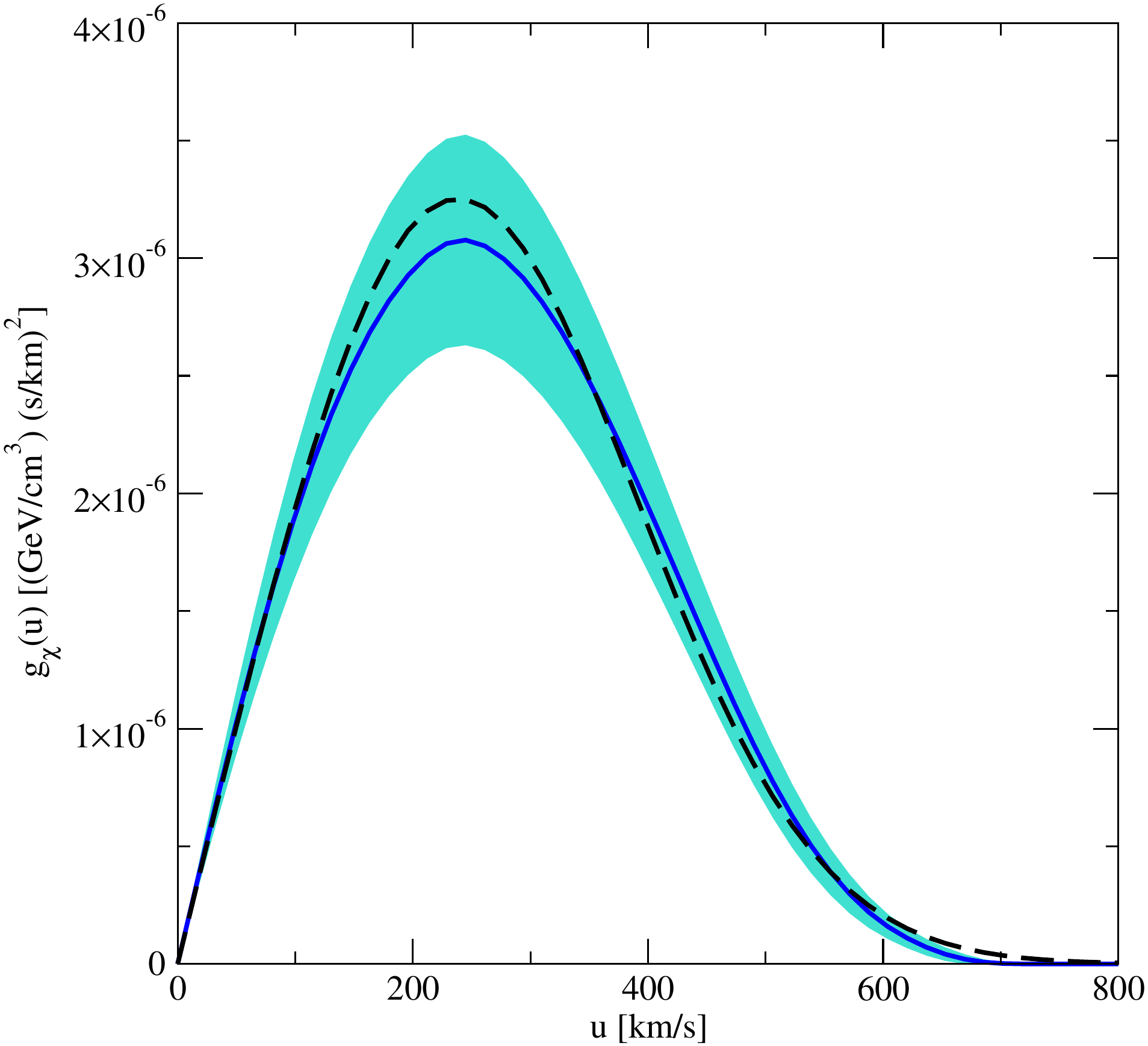}
\end{minipage}
\caption{Dark matter distribution bands ($\pm$ 2 standard deviations from the mean) encoding all the information contained in the dynamical constraints of section~\ref{sec:dyncon}. We show the time averaged local velocity distribution integrated over angles as defined in Eq.~\eqref{eq:g}. The left panel of this figure has been obtained by applying the procedure described in the text to the set of parameters ($w$, $r_a$, $\gamma$) determining the blue dashed curve in the left panel of Fig.~\ref{fig:beta}, while the right panel corresponds to the same analysis performed assuming the set of parameters ($w$, $r_a$, $\gamma$) associated with the red dotted curve in the left panel of Fig.~\ref{fig:beta}. The blue solid curve is the mean dark matter distribution function while the black dashed line corresponds to a  Maxwell-Boltzmann distribution characterized by $\rho_\textrm{loc}=0.3\,\textrm{GeV}\,\textrm{cm}^{-3}$ and $v_c(R_0)=220\,\textrm{km}\,\textrm{s}^{-1}$.}
\label{fig:uDF}
\end{figure}

We now focus on our benchmark distribution function introduced in section~\ref{sec:OM} based on the superposition of a constant-anisotropy part and the Osipkov-Merritt ansatz. This is a function of the galactic model parameters -- determining $\rho$ and $\Psi$ -- and of the velocity $v$ of the dark matter particles. Indeed, in the notation adopted so far $v$ enters both the expression for $\eps$ and the one for $L$. For any given velocity we want to determine the mean value of the local dark matter distribution function evaluated at that velocity and the corresponding standard deviation -- encoding the astrophysical uncertainties discussed in section~\ref{sec:dyncon}. To this aim we proceed as follows: first we introduce a finite set of dark matter velocities. For each velocity of this set we apply Eq.~(\ref{eq:gpdf}) obtaining the 1-dimensional posterior PDF for the distribution function evaluated at that velocity. For each velocity of the set, the mean of this PDF is the value of the ``mean dark matter distribution function'' associated with that dark matter velocity. The standard deviation of the same PDF gives instead an estimate of the uncertainty within which the dark matter distribution function is known at the velocity. Then, applying this procedure to a sufficiently large set of velocities we can construct ``dark matter distribution bands'' (rather than functions) encoding all the information contained in the data, as well as all the sources of uncertainties included in the Likelihood function. 

\begin{figure}[t]
\begin{minipage}[t]{0.32\linewidth}
\centering
\includegraphics[width=\textwidth]{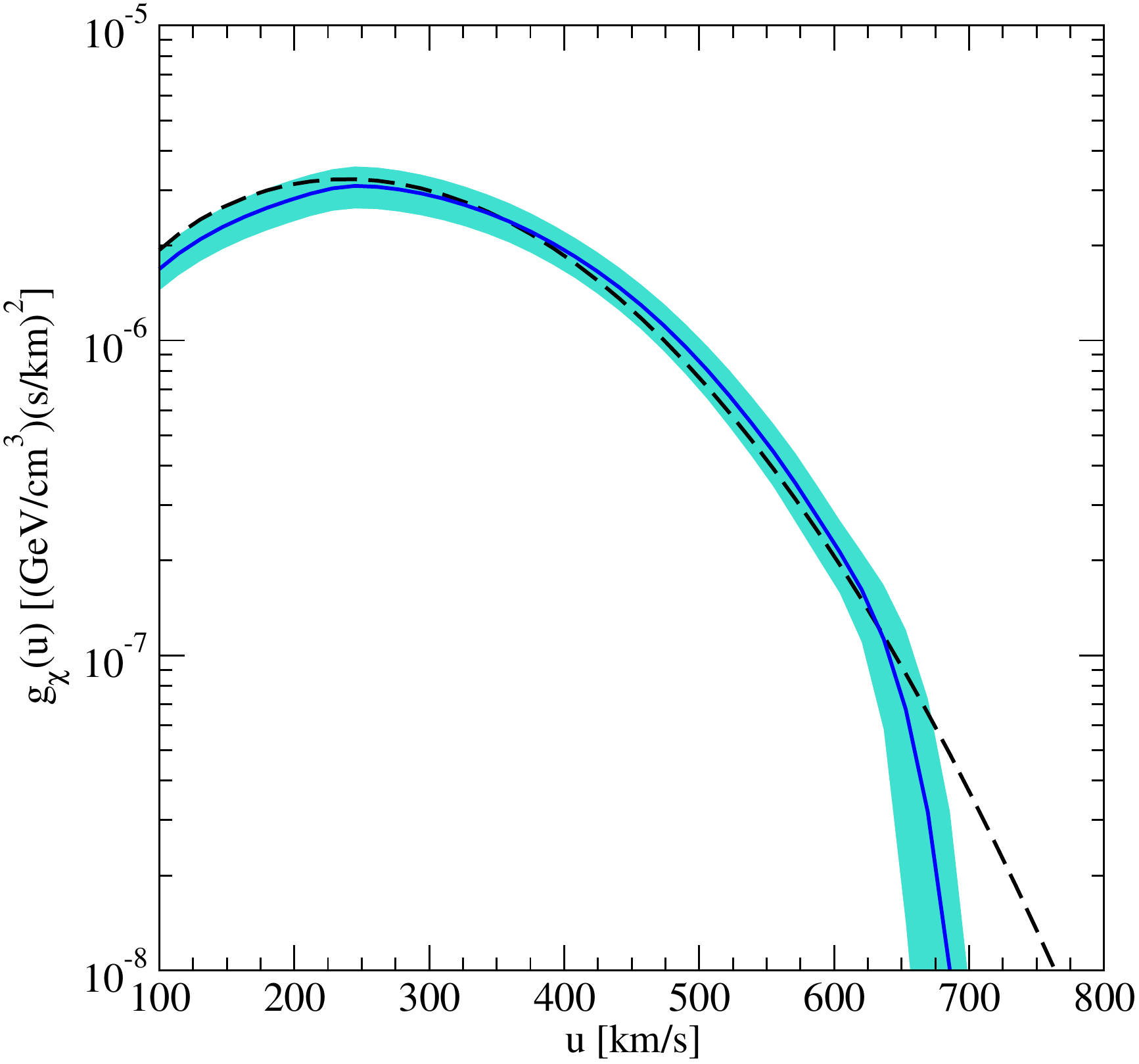}
\end{minipage}
\begin{minipage}[t]{0.32\linewidth}
\centering
\includegraphics[width=\textwidth]{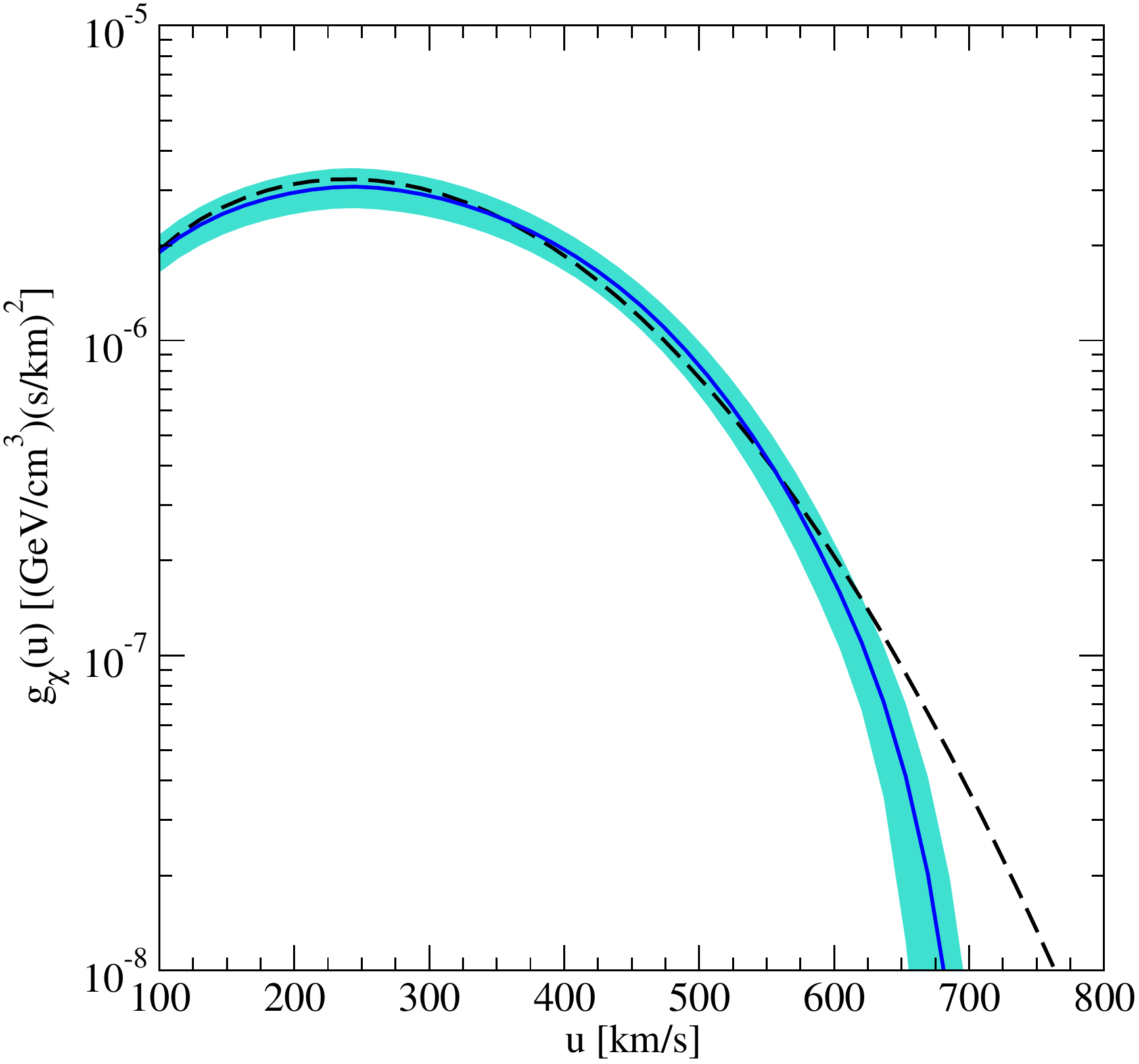}
\end{minipage}
\begin{minipage}[t]{0.327\linewidth}
\centering
\includegraphics[width=\textwidth]{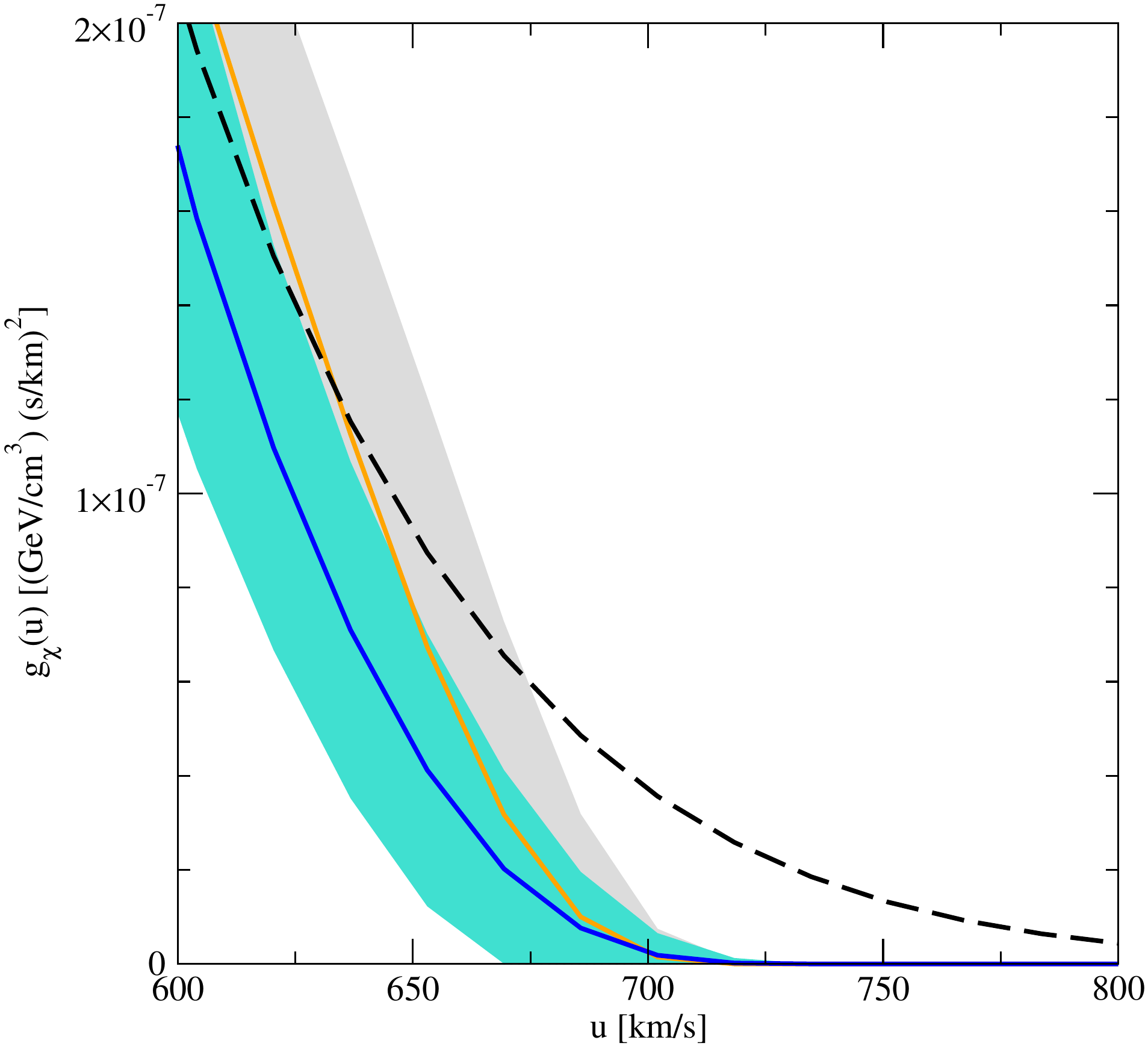}
\end{minipage}
\caption{In the left (central) panel of this figure we compare the tail of the distribution function shown in the left (right) panel of Fig.~\ref{fig:uDF} with the tail of a Maxwell-Boltzmann distribution characterized by $\rho_\textrm{loc}=0.3\,\textrm{GeV}\,\textrm{cm}^{-3}$ and $v_c(R_0)=220\,\textrm{km}\,\textrm{s}^{-1}$. In the right panel, instead, we superimpose in the same plot the tails of the two bands shown in the left and central panels of this figure. The gray band corresponds to the left panel while the turquoise band to the central panel. The blue and orange curves represent the corresponding mean distribution functions while the black dashed line is associated with the reference Maxwell-Boltzmann distribution.}
\label{fig:uDFtail}
\end{figure}

In Fig.~\ref{fig:uDF} we show the time averaged local velocity distribution integrated over angles. This is the distribution relevant for dark matter direct detection experiments and we give the precise definition in Eq.~\eqref{eq:g} below. We show bands for the distribution, encoding the astrophysical uncertainties discussed in section~\ref{sec:dyncon} obtained adding and subtracting two standard deviations to the mean distribution function constructed as explained above. The left panel of this figure has been obtained by applying the procedure described here to the set of parameters ($w$, $r_a$, $\gamma$) determining the blue dashed curve in the left panel of Fig.~\ref{fig:beta}, while the right panel of Fig.~\ref{fig:uDF} corresponds to the same analysis performed assuming the set of parameters ($w$, $r_a$, $\gamma$) associated with the red dotted curve in the left panel of Fig.~\ref{fig:beta}. For comparison, in Fig.~\ref{fig:uDFtail} we superimpose the tails of the two bands shown in the panels of Fig.~\ref{fig:uDF} to the tail of a Maxwell-Boltzmann distribution characterized by $\rho_\textrm{loc}=0.3\,\textrm{GeV}\,\textrm{cm}^{-3}$ and $v_c(R_0)=220\,\textrm{km}\,\textrm{s}^{-1}$ (left and central panels). In the right panel of Fig.~\ref{fig:uDFtail}, instead, we superimpose in the same plot the tails of the two bands shown in the left and central panels of the same figure. Therefore, in this plot the gray band corresponds to the left panel of Fig.~\ref{fig:uDF} while the turquoise band to the right panel of this figure. From the right panel of Fig.~\ref{fig:uDFtail} one can see that in the high velocity tail of the dark matter distribution function, relevant for light WIMP searches, the uncertainties in the galactic model parameters as well as the uncertainties in the dark matter anisotropy parameter can be simultaneously bracketed considering the upper limit of the gray band and the lower limit of the turquoise band. In the next section we will use these distribution functions to discuss the impact of anisotropic dark matter distribution functions on the direct detection of WIMPs.

\section{Dark matter direct detection data}
\label{sec:DD}

\subsection{Event rates}

The differential rate in events/keV/kg/day for a dark matter particle $\chi$ to scatter elastically in a detector composed of nuclei with mass number $A$ and charge  $Z$, and depositing the nuclear recoil energy
$E_R$ is
\beq \label{rate}
\frac{d R}{d E_R} = \frac{\rho_{\rm loc}}{m_\chi} \frac{1}{m_A}\int_{v>\vmin}d^3 v \frac{d\sigma_A}{d{E_R}} v \tilde f_{\rm det}(\vect v, t).
\eeq
Here $m_A$ and $m_\chi$ are the nucleus and dark matter masses, $\sigma_A$ the
dark matter--nucleus scattering cross section and $\vect v$ the 3-vector
relative velocity between $\chi$ and the nucleus, while $v\equiv
|\vect{v}|$.  $\tilde f_{\rm det}(\vect v, t)$ is the dark matter velocity
distribution in the detector rest frame normalized to one. It is
related to the distribution function $f(\eps, L)$ (with $\eps$ and $L$
considered as functions of $r$ and $\vect v$) by
\begin{equation}
  \tilde f_{\rm det}(\vect v, t) = 
  \frac{1}{\rho_{\rm loc}} f(R_0, \vect v + {\vect v}_{\rm Earth}(t)) \,,
\end{equation}
where ${\vect v}_{\rm Earth}(t)$ is the velocity of the Earth relative
to the halo, including the Sun's motion in the Galaxy as well as the
Earth's revolution around the Sun, which introduces the time dependence.
For a dark matter particle to
deposit recoil energy $E_R$ in the detector a minimal velocity $\vmin$ is
required, restricting the integral over velocities in Eq.~\eqref{rate}. For elastic scattering we have
\beq
\vmin =\sqrt{\frac{m_A E_R}{2 \mu_{\chi A}^2}},
\label{v-inelastic}
\eeq
where $\mu_{\chi A}$ is the reduced mass of the dark matter--nucleus system.

The particle physics enters in Eq.~\eqref{rate} through the differential cross section which is in general a sum of spin-independent and spin-dependent contributions. In this paper we consider only spin-independent WIMP interactions, for which the differential cross section is 
\begin{align}
  \frac{d\sigma_A}{dE_R} = \frac{m_A}{2\mu_{\chi A}^2 v^2} \sigma_A^0 F^2(E_R) \,, 
  \label{eq:dsigmadE}
\end{align}
where $\sigma_A^0$ is the total dark matter--nucleus scattering cross section
at zero momentum transfer, and $F(E_R)$ is a form factor. For $F(E_R)$ we use the Helm~\cite{Helm} form factor. The
astrophysics dependence enters in Eq.~\eqref{rate} through the dark matter velocity
distribution $\tilde f_{\rm det} (\vect{v}, t)$ in the detector rest frame. Defining the halo integral
\beq\label{eq:eta} 
\eta(\vmin, t) \equiv \int_{v > \vmin} d^3 v \frac{\tilde f_{\rm det} (\vect{v}, t)}{v} \,,
\eeq
the event rate is given by
\beq\label{eq:Rgamma}
\frac{d R}{d E_R} = \frac{\rho_{\rm loc} \sigma_A^0 F^2(E_R)}{2 m_\chi \mu_{\chi A}^2} \, \eta(\vmin, t) \,.
\eeq
We can also write the halo integral in terms of the dark matter phase-space density function $g_\chi (v,t)$, 
\beq
\eta(\vmin, t) = \frac{1}{\rho_{\rm loc}} \int_{\vmin}^\infty dv \, g_\chi (v,t) \,,
\eeq
where
\beq\label{eq:g}
g_\chi (v,t) = v \rho_{\rm loc} \int d \Omega_{\vect v} \tilde f_{\rm det} (\vect{v}, t)
 =  v \int d \Omega_{\vect v} f(R_0, \vect v + {\vect v}_{\rm Earth}(t)) \,.
\eeq

\subsection{Description of the used data}
\label{DirectDetData}

Let us now discuss the details on how we perform fits to data from various direct detection experiments. We consider the most recent experimental data sets available from the following experiments: DAMA, CDMS, XENON100, XENON10, CoGeNT, CRESST, KIMS and LUX.

\bigskip

{\bf DAMA}:
We use the data on the modulation amplitude for the 1.17 ton yr DAMA exposure given in Fig.~6 of Ref.~\cite{DAMA}, divided into 12 bins. In our fit we use the signal region from 2 keVee to 8 keVee. Above this energy range the data is consistent with no modulation. The signal as a function of energy and time can be written as,
\be
S(E, t) = S_0 (E) + A (E) \cos \omega (t - t_0),
\ee
where $E$ is the measured energy (in keVee), $S_0$ is the unmodulated signal, $A(E)$ is the annual modulation amplitude, $\omega=2 \pi/1$ yr, and $t_0=152$ days. Our analysis of the DAMA data is analogous to those presented in~\cite{Fairbairn:2008, Kopp:2010, Kopp:2011}.

The quenching factor of Na, $q_{\rm Na}$, is an important parameter in the analysis of low mass WIMPs. A recent measurement of $q_{\rm Na}$ shows a decrease of the quenching factor at lower energies~\cite{Collar:2013}. This would result in a shift of the allowed region of DAMA sodium in the cross section versus mass plane towards higher masses, see {\it e.g.,}~\cite{Kopp:2011}. Since $q_{\rm Na}$ is difficult to measure, there are some uncertainties regarding its measured value. In this work, we use $q_{\rm Na}=0.3$ and $q_{\rm I}=0.09$ for the quenching factors of Na and I, respectively, as measured by the DAMA collaboration~\cite{Bernabei:1996}. We consider a 10\% uncertainty in the value of $q_{\rm Na}$ in our fit. According to the results of Ref.~\cite{Bozorgnia:2010xy} the effect of ion channeling in NaI is tiny and therefore we neglect it.

To fit the DAMA data, we construct a $\chi^2$ function
\be
\chi^2_{\rm DAMA} (m_\chi, \sigma_p) = \sum_{i=1}^{i=12} \left(\frac{A_i^{\rm pred} (m_\chi, \sigma_p) - A_i^{\rm obs}}{\sigma_i} \right)^2,
\label{eq: DAMAchisq}
\ee
where $A_i^{\rm obs}$ and $\sigma_i$ are the experimental data points and their errors, respectively, from Fig.~6 of Ref.~\cite{DAMA}. The sum is over the 12 energy bins. The best fit point can be found by minimizing Eq.~(\ref{eq: DAMAchisq}) with respect to the WIMP mass $m_\chi$, and cross section $\sigma_p$. The allowed regions in the mass -- cross section plane at a given CL are obtained by looking for contours $\chi^2 (m_\chi, \sigma_p) =\chi^2_{\rm min} + \Delta \chi^2 ({\rm CL})$, where $\Delta \chi^2 ({\rm CL})$ is evaluated for 2 degrees of freedom (dof), e.g., $\Delta \chi^2 (90\%) = 4.6$ and $\Delta \chi^2 (99.73 \%) = 11.8$.

\bigskip

{\bf CDMS}:
The CDMS-II collaboration has observed two events with recoil energies of 12.3 keV and 15.5 keV in their data taken with Ge detectors with an exposure of 612 kg days in four periods between July 2007 and September 2008~\cite{CDMS:2009}. As done in Ref.~\cite{sz:2011}, we use the maximum gap method from Ref.~\cite{Yellin} to set exclusion limits (labeled ``CDMS--Ge" in Figs.~\ref{fig:ExcLim-Iso}, \ref{fig:ExcLimTS}, \ref{fig:ExcLim}, and \ref{fig:Excl-baryons}). We use a constant energy resolution of 0.2 keV, and take into account a linear efficiency drop from 32\% at 20 keV to 25\% at 10 keV and 100 keV. 

In a modified reanalysis of the CDMS-II data that was collected in eight Ge detectors between October 2006 and September 2008~\cite{CDMS-lt}, the CDMS collaboration lowered the analysis energy threshold to 2 keV allowing for a larger background. In this analysis, they obtained a higher sensitivity to WIMPs with masses lower than $\sim 10$ GeV. We analyze the low-threshold data in a similar manner as in Ref.~\cite{sz:2011}, including only bins with a larger predicted number of events compared to the observed one. The exclusion limit from our analysis is labeled ``CDMS-LT" in Figs.~\ref{fig:ExcLim-Iso}, \ref{fig:ExcLimTS}, \ref{fig:ExcLim}, and \ref{fig:Excl-baryons}.

Recently, the CDMS collaboration presented an analysis of data taken with Si detectors with an exposure of 140.2 kg days using four run periods between July 2007 and September 2008~\cite{CDMS-Si:2013} which revealed 3 events in the dark matter search region. The total estimated background was 0.62 events. Our analysis is similar to the one in Ref.~\cite{Frandsen}. We use the extended maximum likelihood method~\cite{Barlow} to calculate the allowed parameter region (labeled ``CDMS--Si" in Figs.~\ref{fig:ExcLim-Iso}, \ref{fig:ExcLimTS}, \ref{fig:ExcLim}, and \ref{fig:Excl-baryons}). To include the background, we rescale the individual background contributions from Ref.~\cite{McCarthy}, such that 0.41, 0.13, and 0.08 events are expected from surface events, neutrons, and $^{206}$Pb, respectively. We use the detector acceptance from Ref.~\cite{CDMS-Si:2013} and assume an energy resolution of 0.3 keV. 

\bigskip

{\bf XENON100}:
The XENON100 experiment uses liquid xenon and measures both ionization and scintillation signals. In the 224.6 live days $\times$ 34 kg exposure of XENON100, the two candidate events observed are consistent with the background expectation of $(1.0 \pm 0.2)$ events, and therefore there is no evidence for dark matter interactions~\cite{XENON100}. We analyze the data and derive an exclusion limit using the maximum gap method~\cite{Yellin}. The scintillation light yield, $L_{\rm eff} (E_R)$ is one of the important inputs to the exclusion limits from XENON100. For $L_{\rm eff} (E_R)$ we use the the black solid line from Fig.~1 of \cite{XENON100:2011}. In our analysis we take into account upward fluctuations due to Poisson statistics of events below the threshold, which is important for the low WIMP mass region.

\bigskip

{\bf XENON10}:
We use the S2 analysis for the XENON10 experiment~\cite{XENON10}, using results from a 12.5 live day dark matter search obtained between 23 August and 14 September, 2006. For the ionization yield, we use the choice of $\mathcal{Q}_y$ made in Ref.~\cite{XENON10} assuming that it would vanish for $E_R<1.4$ keV. To derive an exclusion limit we use the maximum gap method. Our results are consistent with Ref.~\cite{Frandsen}.

\bigskip

{\bf CoGeNT}:
The CoGeNT experiment uses very low threshold germanium detectors. We fit the unmodulated CoGeNT data shown in the in--set of Fig.~1 of Ref.~\cite{CoGeNT:2011}. This is the exponential-like irreducible background of events in the bulk of the crystal, after subtracting the L--shell peaks and a constant spectral component. In the in--set of Fig.~1 of Ref.~\cite{CoGeNT:2011}, black and white data points correspond to two different peak-subtraction methods. Similar to Ref.~\cite{sz:2011}, we derive our fit by taking the average of the black and white data points, and to be conservative we use the lowest and highest edges of the error bars to account for this systematic uncertainty in the fit. We assume that the total excess events are explained by dark matter. Let us mention, however, that those events might be contaminated by background activity on the surface of the detector (``surface events'') \cite{CoGeNT:TAUP2011}, which will significantly affect the size and location of the allowed region, see {\it e.g.}~\cite{Kopp:2011}.

The CoGeNT collaboration has also reported an annual modulation signal at low energies~\cite{CoGeNT:2011}, and more recently confirmed that the annual modulation persists in 3.4 yr of data acquired~\cite{CoGeNT:TAUP2013}. The significance of the singal is weak (slightly above $2\sigma$), and therefore, we do not use the CoGeNT data on the annual modulation amplitude in this paper, see {\it e.g.}~\cite{HerreroGarcia:2011aa, sz:2011} for a discussion.

\bigskip

{\bf CRESST}:
The CRESST-II experiment uses CaWO$_4$ crystals and has completed 730 kg days of data taking~\cite{CRESST}. They find 67 events in the acceptance region where a dark matter signal is expected. To fit the data from CRESST, we use a method analogous to the one used in Ref.~\cite{Kopp:2011}. In particular, we use publicly available information to fit the total event rate in each detector module and the overall energy spectrum, without including the light yield for each event.

\bigskip

{\bf KIMS}:
The KIMS experiment uses an array of 12 CsI scintillators to search for WIMPs. We use the most recent KIMS result based on an exposure of 24524.3 kg days~\cite{Kim:2012rza}. KIMS does not see any events at recoil energies below 8 KeVee and so they exclude the possibility of explaining the DAMA annual modulation by dark matter particles recoiling on iodine.

\bigskip

{\bf LUX}:
The LUX (Large Underground Xenon) experiment has just released its first results~\cite{LUX:2013}. In their analysis of 85.3 live-days of data taken in the period of April to August 2013, the data is consistent with the background-only hypothesis. The collaboration sets a 90\% confidence limit on the spin-independent elastic WIMP-nucleon cross section, with a minimum upper limit of $7.6 \times 10^{-46}$ cm$^2$ on the cross section for a WIMP mass of 33 GeV assuming the Standard Halo Model. The LUX results are in strong disagreement with signals from the DAMA, CoGeNT, CRESST, and CDMS-Si experiments if the Standard Halo Model is assumed. With the available information it is not possible to reproduce the  likelihood analysis performed by the LUX collaboration. Therefore we  employ the maximum gap method to set an upper limit on the cross  section. We consider as signal region the region below the mean of the 
Gaussian  fit to the nuclear recoil calibration events (red solid curve in Fig.~4 of \cite{LUX:2013}) and assume an acceptance of 0.5. It can be seen from Fig.~4 of \cite{LUX:2013} that one event at 3.1 photoelectrons falls on the red solid curve, thus we can consider either zero or one event in our analysis. To find the relation between S1 and nuclear recoil energy $E_R$, we use Fig.~4 of \cite{LUX:2013} and find the value of S1 at the intersection of the mean nuclear recoil curve and each recoil energy contour. For the efficiency as a function of recoil energy, we interpolate the black points in Fig.~9 of \cite{LUX:2013} for events with a corrected S1 between 2 and 30 photoelectrons and a S2 signal larger than 200 photoelectrons. We multiply the efficiency from Fig.~9 of \cite{LUX:2013}  by 0.5 to find the total efficiency for our maximum gap method, and set it equal to zero below $E_R=3$ keV. Assuming the Standard Halo Model with the Maxwellian velocity distribution and parameters chosen as in \cite{LUX:2013}, we find that our 90\% CL contour agrees with good 
accuracy with the limit set by the LUX collaboration if we assume zero events. On the other hand, if we assume one event makes the cut, our limit would not match as closely the limit set by the LUX collaboration. Therefore in this work we will consider zero events when deriving exclusion limits.

\bigskip

Finally, we would like to mention the very recent results from the CDMSlite and MALBEK detectors which search for light WIMPs, although we do not perform fits to their data in this work. CDMSlite~\cite{CDMSlite:2013} is a calorimetric technique used by the SuperCDMS experiment that substantially reduces the energy threshold and improves the energy resolution, resulting in a significantly better sensitivity to light WIMPs with masses $<10$ GeV. In the recent analysis by CDMSlite, with an exposure of 6 kg days and without any background subtraction, new limits on the WIMP mass and cross section were obtained for WIMPs of mass $<6$ GeV, excluding parts of CDMS--Si and CoGeNT allowed regions. 

The Majorana demonstrator~\cite{MALBEK:TAUP13} uses an array of high purity Ge detectors with sub-keV energy threshold  to search for light WIMPs with masses $<10$ GeV. Recently, limits at the 90\% confidence in the WIMP parameter space were presented from 221 day dataset obtained by MALBEK (Majorana Low-background BEGe Detector at Kurf). The MALBEK exclusion limits also rule out parts of CDMS--Si and CoGeNT preferred regions. 

\section{Results}
\label{sec:results}

\begin{figure}
\centering
\includegraphics[height=210pt]{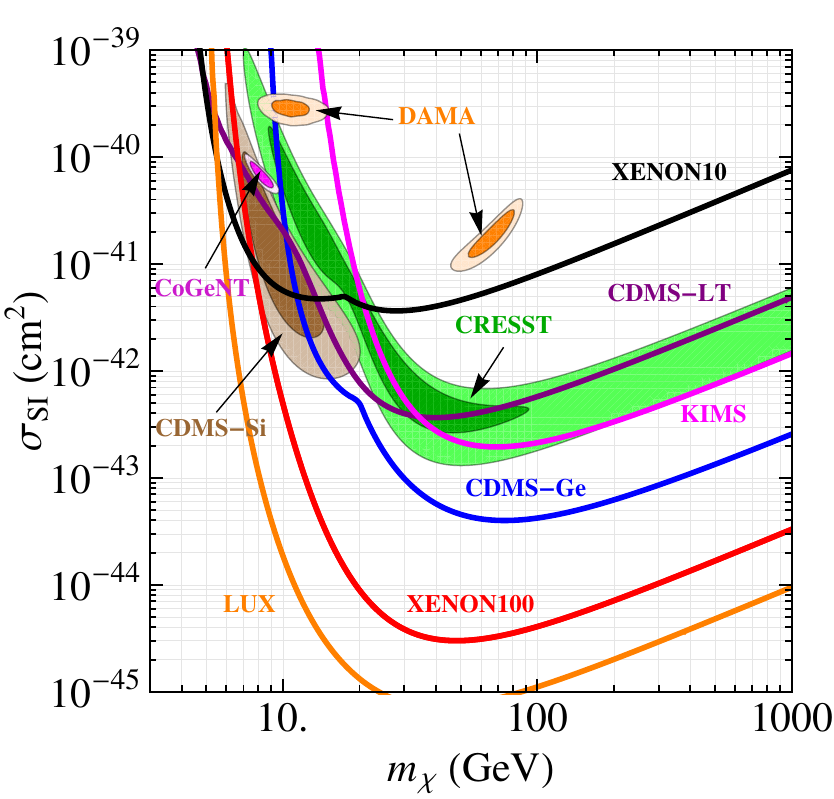}
\includegraphics[height=210pt]{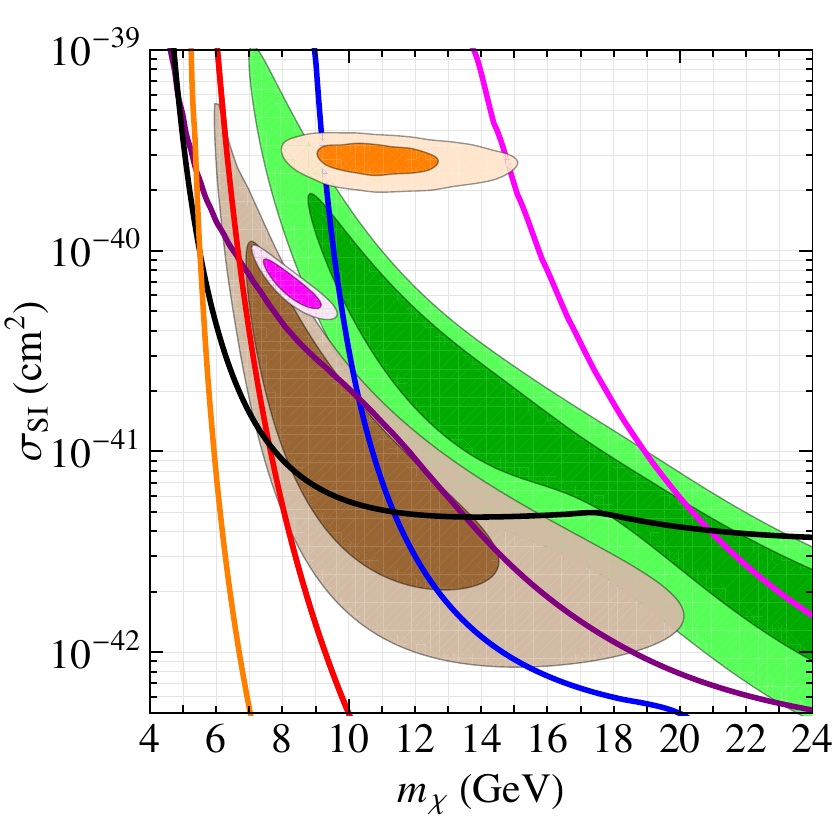}\\
\caption{Constraints on elastic, spin-independent dark matter--nucleon scattering for an isotropic velocity distribution. The preferred regions of DAMA, CoGeNT, and CRESST (at 90\% CL and 3$\sigma$), and CDMS--Si (at 68\% and 90\% CL) are shown together with constrains from XENON100, XENON10, CDMS--Ge, CDMS--LT, KIMS, and LUX (at 90\%~CL). The left panel shows the results for a wide range of dark matter masses, while the right panel zooms on the low mass region (same color code as in the left panel).}
\label{fig:ExcLim-Iso}
\end{figure}

We now move to the analysis of direct detection data in the light of the halo models discussed above. As a reference point, we consider first an isotropic dark matter velocity distribution. We adopt the best fit model for the galaxy from the analysis of kinematical data and perform the Eddington inversion to calculate the velocity distribution (see dashed curves in Fig.~\ref{fig:uDF-cases} below). Fig.~\ref{fig:ExcLim-Iso} shows the corresponding exclusion limits and allowed regions in the plane of dark matter mass and spin-independent cross section from the experiments discussed in Section~\ref{DirectDetData}. The zoom to the low-mass region in the right pannel illustrates the well known fact that the hints for a positive signal from DAMA, CoGeNT, CRESST, and CDMS--Si are in tension with limits from XENON10, XENON100, CDMS--Ge, CDMS--LT, and LUX. We now revisit this problem considering the anisotropic halo models discussed above, as well as the uncertainties from the fit to kinematical Milky Way data.

\subsection{The impact of anisotropy}

\begin{figure}
\centering
\includegraphics[height=130pt]{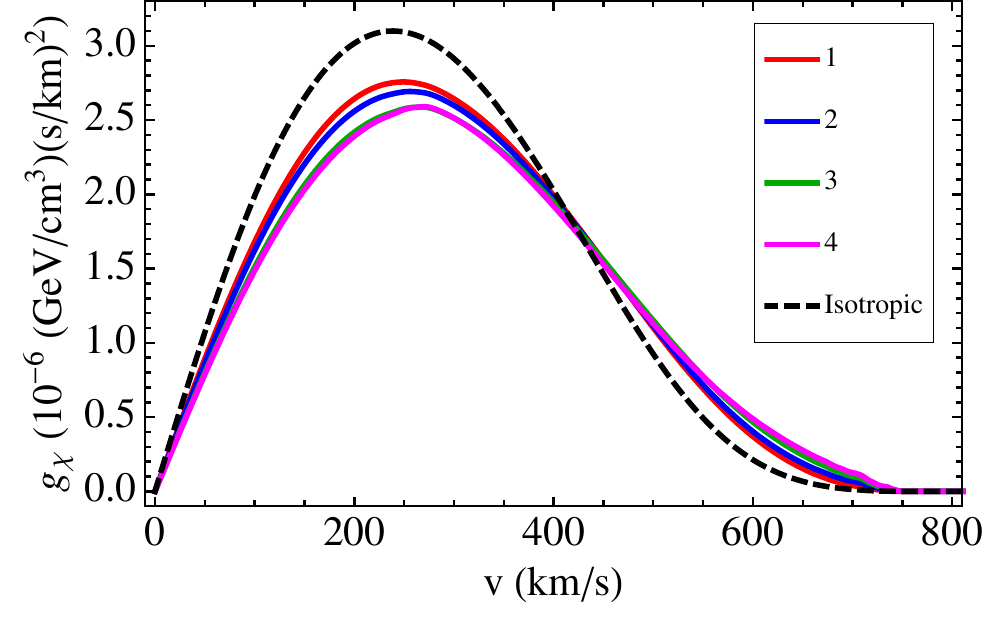}
\raisebox{0mm}{\includegraphics[height=140pt]{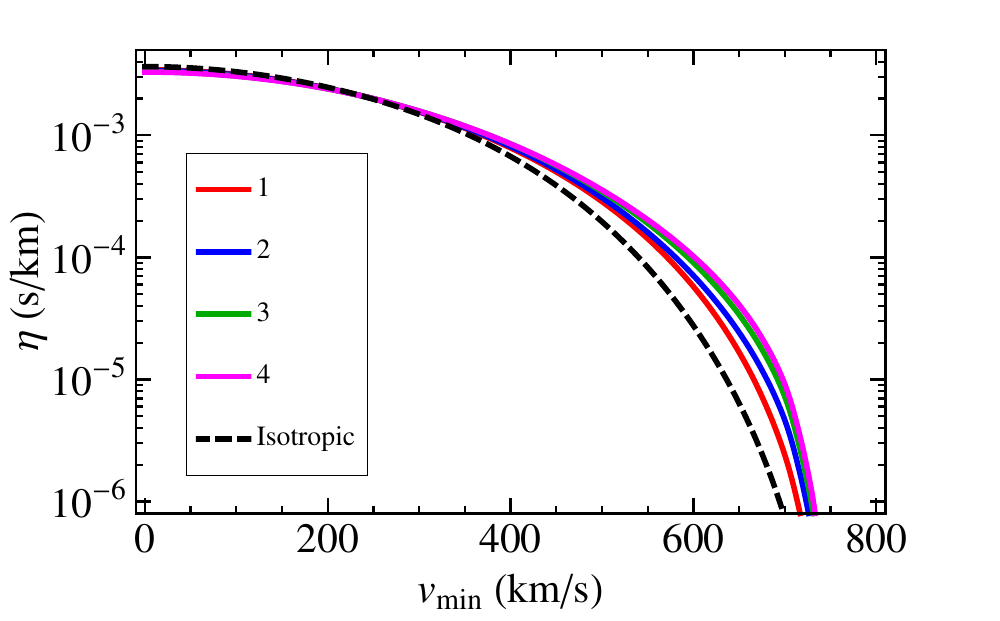}}
\caption{Velocity distributions for the four halo models discussed in section~\ref{sec:alt-models} as defined in Tab.~\ref{tab:cases}. The left pannel shows the time averaged local velocity distribution integrated over angles defined in Eq.~\eqref{eq:g}. The right panel shows the halo integral $\eta$ as a function of the minimial velocity defined in Eq.~\eqref{eq:eta}. The dashed curve in both pannels corresponds to an isotropic velocity distribution. All curves are based on the same model for the spatial visible and dark mass distributions.}
\label{fig:uDF-cases}
\end{figure}

\begin{figure}
\centering
\includegraphics[height=210pt]{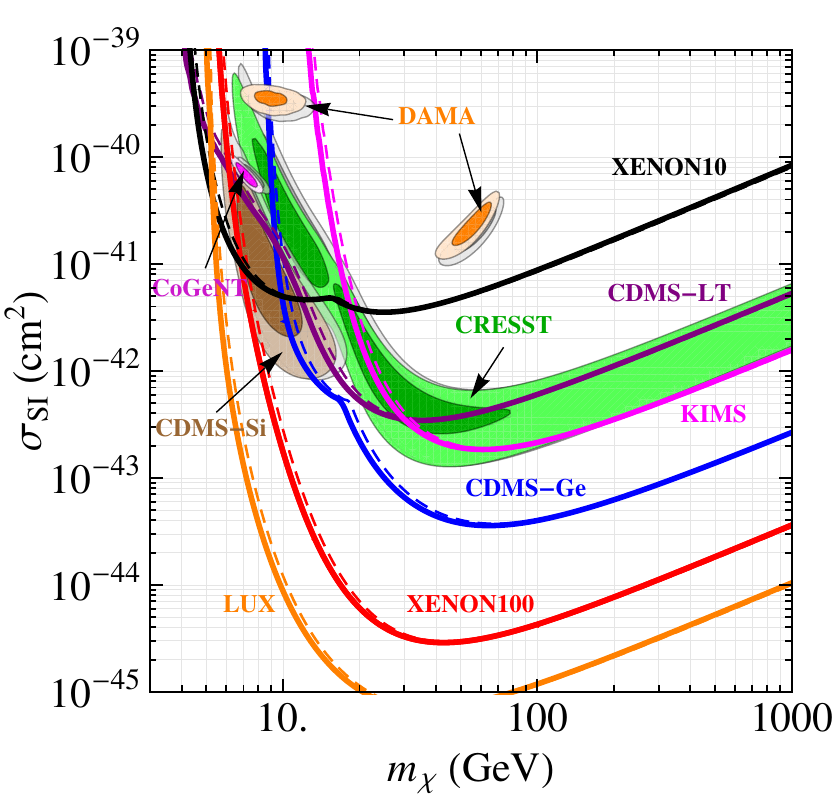}
\includegraphics[height=210pt]{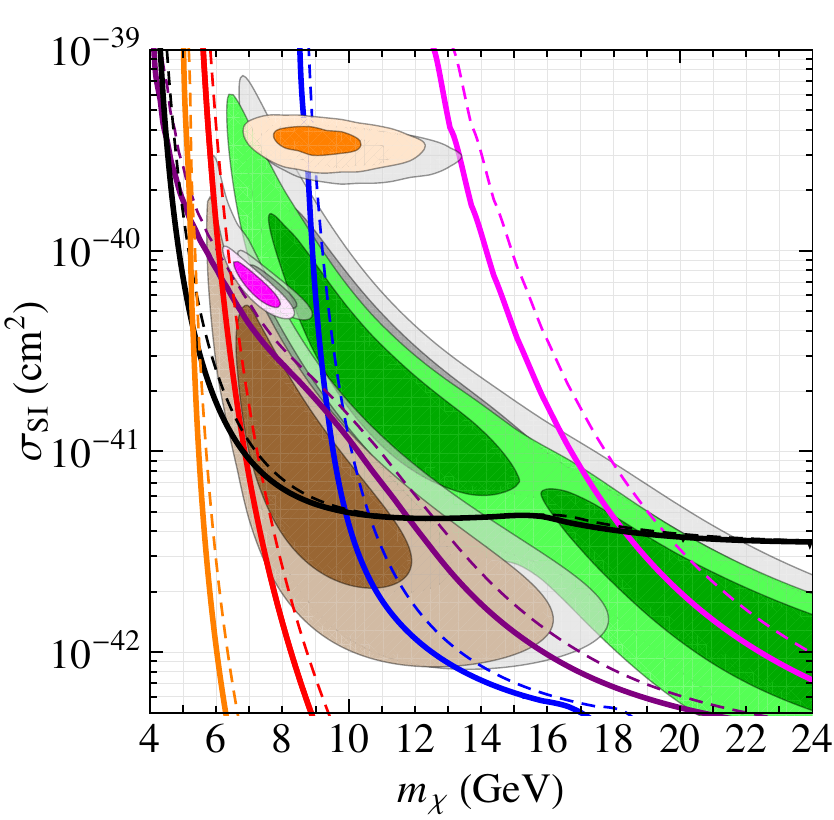}\\
\includegraphics[height=210pt]{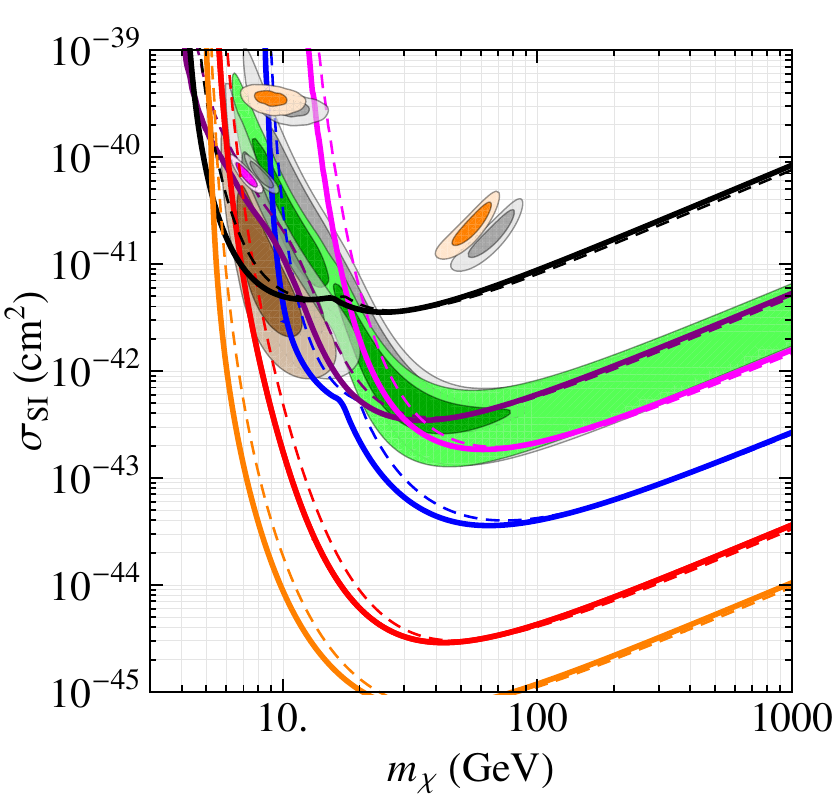}
\includegraphics[height=210pt]{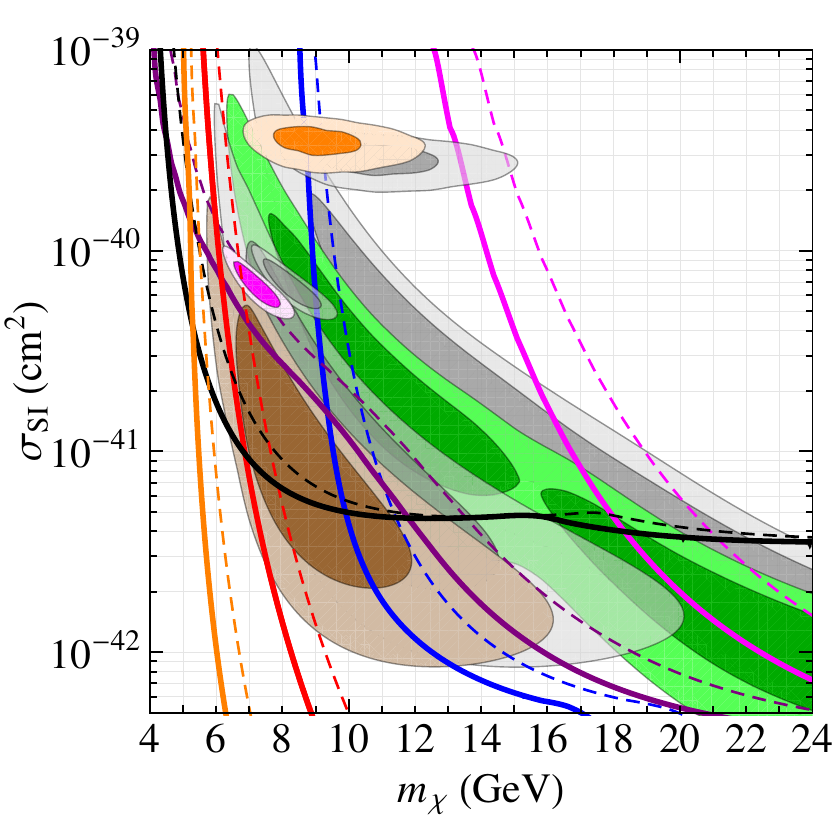}\\
\caption{Same as Fig.~\ref{fig:ExcLim-Iso}, but using anisotropic halo models defined in Tab.~\ref{tab:cases}. We compare  ``Model 4'' with ``Model 1'' in the upper panels, and ``Model 4'' with the isotropic case in the lower panels. The exclusion limits and allowed regions obtained for ``Model 4''  are shown by solid lines and in color, whereas the dashed exclusion curves and the gray regions correspond to ``Model 1'' in the upper panels and to the isotropic case in the lower panels.}
\label{fig:ExcLimTS}
\end{figure}

First we keep the model for the Milky Way fixed but allow for anisotropic velocity distributions, considering the models discussed in section~\ref{sec:alt-models}. The distribution function $g_\chi(v)$ as well as the halo integral $\eta(\vmin)$ relevant for direct detection resulting from the 4 anisotropic halo models defined in Tab.~\ref{tab:cases} are shown in Fig.~\ref{fig:uDF-cases}. Noting that the models labeled from 1 to 4 represent increasing anisotropy parameters $\beta$ at large radii (see Fig.~\ref{fig:beta2}) we observe that more radial distributions tend to shift the local velocity distribution to higher velocities. All 4 models have a very similar value of $\beta$ at the galactocentric distance corresponding to the location of the Sun of $\beta(R_0) \approx 0.2$, whereas they differ most significantly at radii $r \sim 10 r_{-2}$, where for the specific example considered here $r_{-2} \approx 18$~kpc and $R_{\rm vir} \approx 270$~kpc. We conclude that the degree of anisotropy at radii of order up to the virial radius has significant impact on the local velocity distribution at our position in the Milky Way. Unfortunately for large radii the shape of $\beta(r)$ from N-body simulations has a very wide range (see {\it e.g.}\ Fig.~3 of Ref.~\cite{Ludlow:2011cs}), indicating a large variability and dependence on the specific merger history of the halo.

Fig.~\ref{fig:ExcLimTS} shows the effect of the anisotropic velocity distributions on the allowed regions and exclusion limits from the experiments. In the upper pannels we compare the Model with the highest  (``4'') to the one with the lowest (``1'') anisotropy, whereas the lower pannels compare Model~4 with the isotropic case (same as shown in Fig.~\ref{fig:ExcLim-Iso}). We observe that the anisotropy affects mainly the low WIMP mass region where experiments probe the high-velocity tail of the distribution. Increasing $\beta$ from the isotropic case to Model 4 shifts the regions to smaller WIMP masses by about 1--2~GeV, since more particles appear in the high-velocity tail. For large WIMP masses ($m_\chi \gtrsim 40$~GeV) limits become insensitive to the anisotropy, since the event rates are dominated by the region $\vmin \simeq 400$~km/s, where the halo integrals $\eta(\vmin)$ become very similar, compare with Fig.~\ref{fig:uDF-cases}. Let us note, however, that regions and limits shift in the same way, and hence the tension between them remains essentially unchanged. We have also checked that the phase of the annual modulation signal is basically unaffected by the anisotropic models considered here and remains at day 152 (June 2nd) as in the case of an isotropic distribution.

\subsection{Astrophysical uncertainties from the fit to Milky Way data}

Let us now investigate the impact on the allowed regions and exclusion limits from taking into account the variations of the parameters for the Milky Way mass model as allowed by the fit to the kinematical data. As discussed in section~\ref{sec:bayes} we bracket the astrophysical uncertainties by considering $(a)$ the upper $2\sigma$ limit from the distribution shown in the left panel of Fig.~\ref{fig:uDF} based on a relatively large anisotropy parameter $\beta(r)$, and $(b)$ the lower $2\sigma$ limit from the distribution in the right panel of Fig.~\ref{fig:uDF} based on a less anisotropic velocity distribution. As argued in section~\ref{sec:bayes} and illustrated in the right panel of Fig.~\ref{fig:uDFtail} those two choices cover the allowed spread in the high-velocity tail of the local dark matter distribution, which is most relevant for the direct detection data in the $m_\chi \sim 10$~GeV region.

\begin{figure}
\centering
\includegraphics[height=210pt]{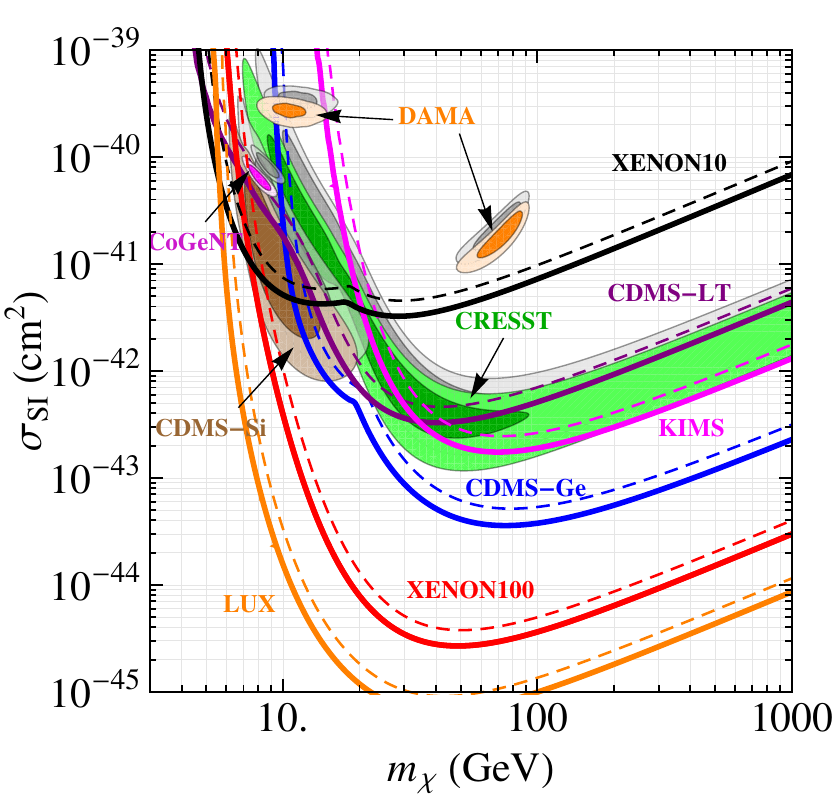}
\includegraphics[height=210pt]{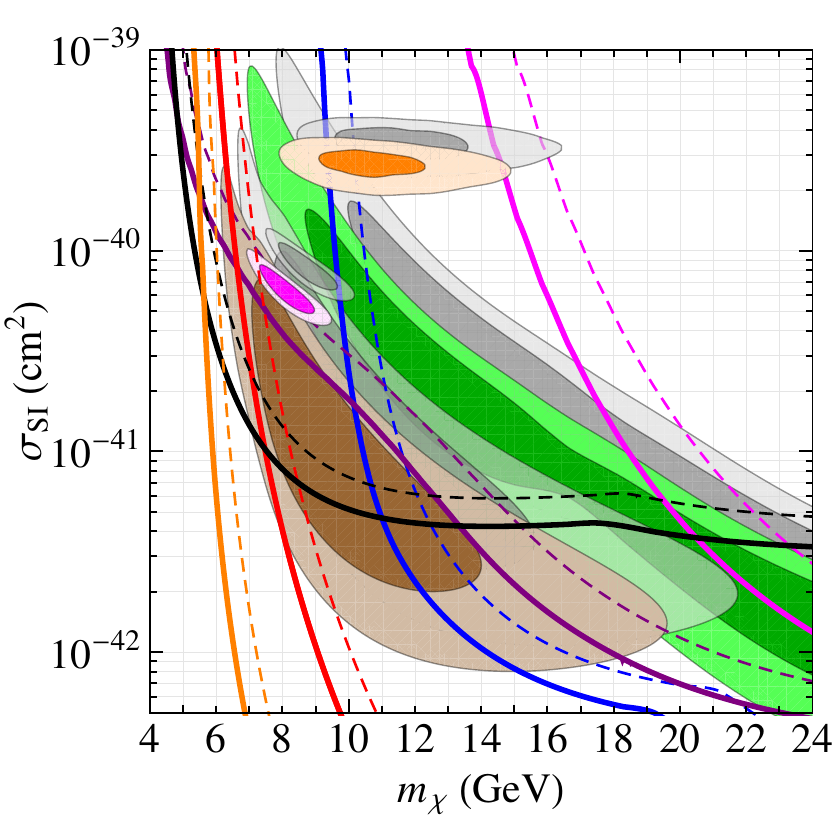}
\caption{Same as Fig~\ref{fig:ExcLim-Iso}, but comparing two anisotropic models including variations of the parameters for the Milky Way mass model. Solid exclusion limits and colored regions correspond to the upper $2\sigma$ limit from the distribution shown in the left panel of Fig.~\ref{fig:uDF} (based on the blue dashed curve for $\beta(r)$ in Fig.~\ref{fig:beta} left), whereas the dashed exclusion limits and gray shaded regions correspond to the lower $2\sigma$ limit from the distribution in the right panel of Fig.~\ref{fig:uDF} (based on the red dotted curve for $\beta(r)$ in Fig.~\ref{fig:beta} left).}
\label{fig:ExcLim}
\end{figure}

The impact on the allowed regions and exclusion limits of changing between these two models is shown in Fig.~\ref{fig:ExcLim}. From the zoom to the low WIMP mass region in the right panel we observe a shift in $m_\chi$ of about 1--2~GeV, similar to the one found in Fig.~\ref{fig:ExcLimTS} where the Milky Way mass model has been kept fixed. However, different from Fig.~\ref{fig:ExcLimTS} we see in the left panel of Fig.~\ref{fig:ExcLim} also a shift of the regions for large WIMP masses. This comes from the effect of changing the parameters of the Milky Way model within their allowed ranges from the fit. In particular, an important effect here is the overall normalization of the distribution, {\it i.e.}\ the uncertainty in the local dark matter density $\rho_{\rm loc}$, which varies within the 95\% credible interval $[0.22,0.36]\,\textrm{GeV}\,\textrm{cm}^{-3}$. Note that changing $\rho_{\rm loc}$ results in an overall vertical shift in the exclusion curves and allowed regions which is the same for all WIMP masses. In Fig.~\ref{fig:ExcLim} again we observe that exclusion limits and allowed regions shift in a similar way, such that the compatibility of them remains basically the same.

\subsection{Effect of baryons}

As a side remark let us mention here the importance of including the visible (baryonic) components of the Milky Way in the analysis. Note that our model for the dark matter halo -- the density profile $\rho(r)$ as well as the anisotropy parameter $\beta(r)$ -- is inspired by pure dark matter N-body simulations, which typically do not include baryonic components. On the other hand, the visible components of the Milky Way as described in section~\ref{sec:MW-model} are essential for fitting the kinematical data and in turn constraining the parameters of the dark matter halo.

\begin{figure}
\centering
\includegraphics[height=190pt]{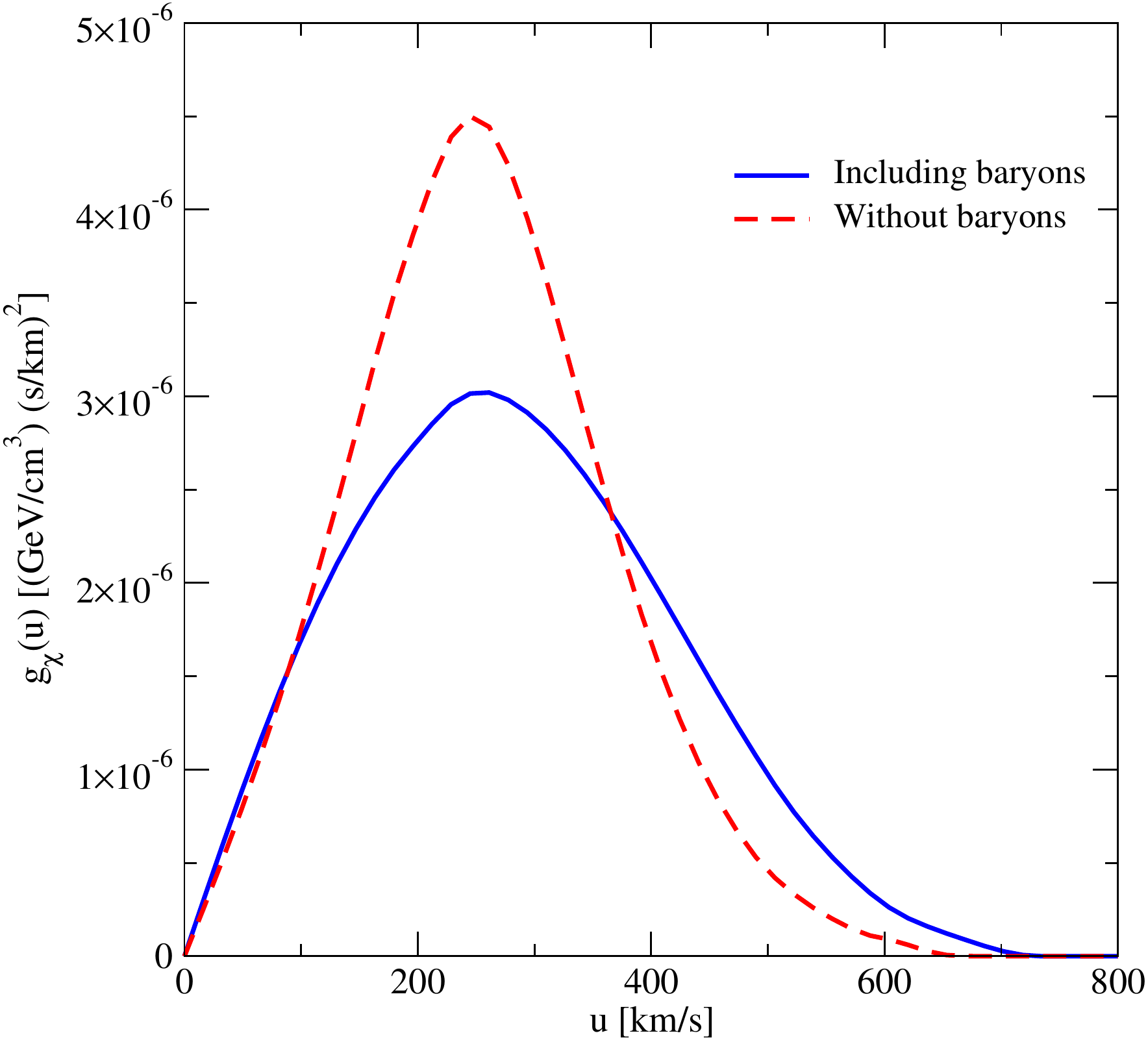}
\includegraphics[height=190pt]{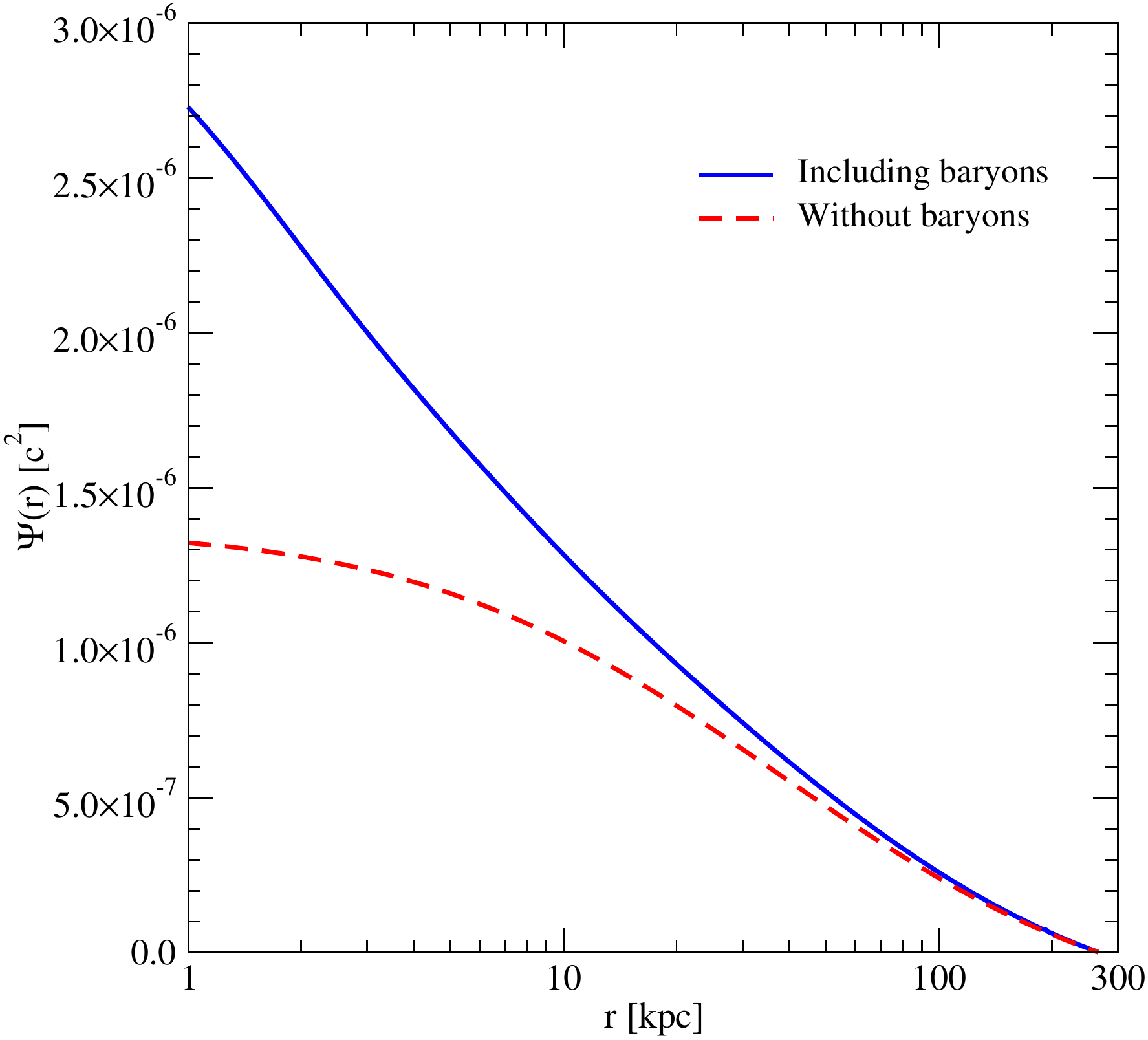}
\caption{Comparison of a dark matter only halo (dashed) and the full mass model of the Milky Way including baryons (solid). The left panel shows the time averaged local velocity distribution integrated over angles defined in Eq.~\eqref{eq:g} and the right panel shows the relative gravitational potential $\Psi(r)$, normalized to zero at the virial radius.}
\label{fig:baryons}
\end{figure}

\begin{figure}
\centering
\includegraphics[height=210pt]{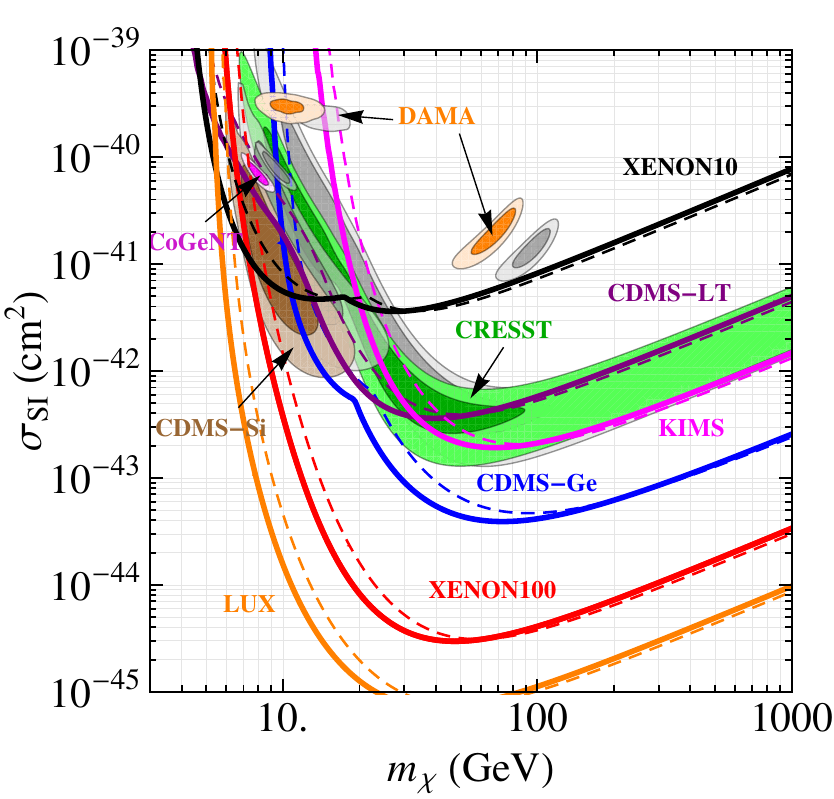}
\includegraphics[height=210pt]{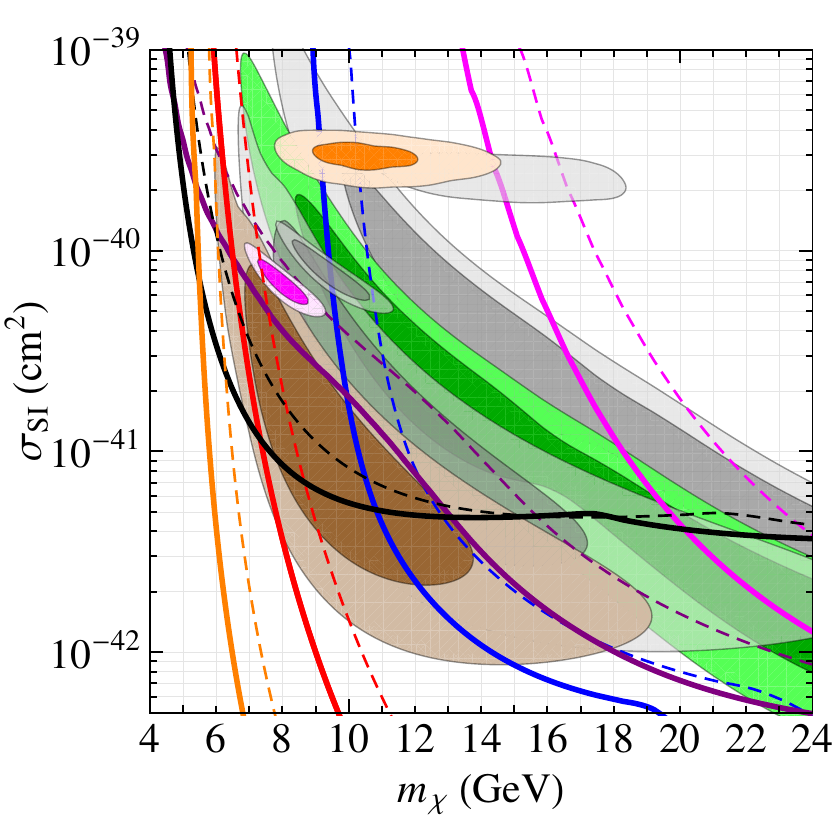}
\caption{Same as Fig~\ref{fig:ExcLim-Iso}, but comparing our standard Milky Way mass model including baryons (solid curves and colored regions) to the same dark matter halo but without baryonic components (dashed curves and gray shaded regions).}
\label{fig:Excl-baryons}
\end{figure}

In Fig.~\ref{fig:baryons} we show the velocity distribution $g_\chi$ as well as the relative potential $\Psi(r)$, where the blue solid curves correspond to our standard best fit model of the Milky Way. For the red dashed curves we use the same dark matter halo as for the solid curves (based on the Einasto profile) but remove all the baryonic components. We see from the right pannel that the baryonic components contribute significantly to the gravitational potential for $r \lesssim 10$~kpc. The left pannel illustrates the impact of baryons on the velocity distribution. It is intuitively clear that the larger gravitational potential increases the number of dark matter particles with high velocities (as well as the escape velocity), a trend which is obvious from the figure. Note that both cases have the same value of the local dark matter density $\rho_{\rm loc}$. While the radial and transversal velocity dispersions differ significantly, the halos with and without baryons lead actually to a rather similar $\beta(r)$ (we explicitly checked this using the anisotropy model characterized by $w=0.15$, $r_a=20$ kpc, $\gamma=-0.10$). 

Fig.~\ref{fig:Excl-baryons} shows the effect of the baryons on the interpretation of direct detection data. Again we find that the low WIMP mass regions is affected by a shift of about 2~GeV, where again the main effect is the larger population of the high-velocity tail of the distribution due to the presence of baryons. Note that for the dark matter only halo the DAMA region around 15~GeV (scattering on Na) appears only at $3\sigma$ with no allowed region at 90\%~CL. The reason is that we draw the allowed regions with respect to the global $\chi^2$ minimum which in this case happens for scattering on iodine ($m_\chi \sim 100$~GeV) with a slightly larger $\Delta\chi^2$ than in the other cases considered before. 

\section{Conclusions}
\label{sec:conclusions}

In this work we have investigated the impact of anisotropic dark matter velocity distributions for direct detection data. We depart from a mass model for the Milky Way including a parameterization of the visible components as well as the dark matter halo, determining the parameters of the model by a detailed fit to kinematical data from the Milky Way. Then we assume a radial profile for the anisotropy parameter $\beta(r)$ motivated by N-body simulations, with a close to isotropic velocity distribution at the center of the galaxy and moderately radial biased distributions at large radii. Self-consistent dark matter distribution functions are derived from the dark matter mass profile $\rho(r)$ and the total gravitational potential $\Psi(r)$ by a generalization of the Eddington inversion procedure to anisotropic velocity distributions. We have investigated the implications for dark matter direct detection by considering the allowed regions and exclusion limits from current data, focusing on spin-independent elastic scattering. Our main findings can be summarized as follows:
\begin{itemize}
\item The local velocity distribution is affected by the degree of anisotropy at radii up to the virial radius.

\item Radially biased velocity distributions at large galactocentric distances lead to an increased high velocity tail of the local dark matter distribution.

\item This leads to a shift of direct detection allowed regions and exlusion limits for WIMP masses around 10~GeV of about 2~GeV, since in this region the high velocity tail is sampled.

\item Exclusion limits for WIMP masses $m_\chi \gtrsim 50$~GeV are less affected by halo anisotropy.

\item Once the full uncertainties from the fit of our Milky Way model are taken into account also the high WIMP mass limits are affected.

\item In general exclusion limits (XENON10, XENON100, CDMS--Ge, CDMS--LT, KIMS, LUX) and allowed regions (DAMA, CoGeNT, CRESST, CDMS--Si) shift in the same way, and the compatibility cannot be improved. 

\item We have shown that the baryonic components of the Milky Way play an important role to determine the local velocity distribution and cannot be neglected when building self-consistent models for the dark matter halo.
\end{itemize}

\subsection*{Acknowledgements}
We would like to thank Mattia Fornasa for many valuable discussions on this subject and for sharing with us the preliminary results of his research on a similar topic. We would also like to thank Piero Ullio for reading a draft version of this work and for his comments and suggestions. Finally, this work has also benefited of several stimulating discussions with Felix Kahlhoefer and Julien Billard at TAUP 2013. We acknowledge support from the  European Union FP7  ITN INVISIBLES (Marie Curie Actions, PITN-GA-2011-289442).

\appendix
\section{Computing the anisotropy parameter}

In this appendix we briefly review the calculation of the anisotropy parameter $\beta(r)$ for distribution functions of the form discussed in this work. Let us first focus on the model discussed in section~\ref{sec:OM} based on the superposition of constant-$\beta$ and Osipkov-Merritt distributions.
We consider here a generalized distribution function given by
\beq
f(\eps,L)= \tilde{G}(Q) L^{2 \gamma}\,;\qquad Q=\eps-\frac{L^2}{2r_a^2}\,.
\eeq
In the limit $\gamma\rightarrow 0$ this expression coincides with the Osipkov-Merritt distribution function, while for $r_a\rightarrow +\infty$ it converges to the case of a distribution function with constant $\beta(r)$. Calculating the anisotropy parameter for this distribution we can thus simultaneously justify all the formulas for $\beta(r)$ given in section~\ref{sec:OM}. First we need to compute the radial and tangential velocity dispersions. For the radial velocity disperison $\sigma_r$ we find
\beqra
\rho\sigma_r^2(r) &=& \pi 2^{\gamma+5/2} \frac{\Gamma(\gamma+1)\Gamma(3/2)}{\Gamma(\gamma+5/2)} \nonumber\\
&\times& \frac{r^{2\gamma}}{(1+r^2/r_a^2)^{\gamma+1}}\int_0^{\Psi(r)} dQ\,\tilde{G}(Q) [\Psi(r) -Q]^{\gamma+3/2}\,,
\label{eq:sigmar}
\eeqra
where we used the identity 
\beq
\int_0^{\pi/2} d\theta\,\sin^{2\gamma+1}\theta\cos^2\theta = \frac{\Gamma(\gamma+1)\Gamma(3/2)}{2\Gamma(\gamma+5/2)}\,,
\eeq
while for the tangential velocity dispersion ($\sigma^2_\theta=\sigma^2_t/2$) one similarly obtains
\beqra
\rho\sigma_\theta^2(r) &=& \pi 2^{\gamma+5/2} \frac{\Gamma(\gamma+2)\Gamma(3/2)}{\Gamma(\gamma+5/2)} \nonumber\\
&\times& \frac{r^{2\gamma}}{(1+r^2/r_a^2)^{\gamma+2}}\int_0^{\Psi(r)} dQ\,\tilde{G}(Q) [\Psi(r) -Q]^{\gamma+3/2}\,,
\label{eq:sigmatheta}
\eeqra
where we used the identity 
\beq
\int_0^{\pi/2} d\theta\,\sin^{2\gamma+3}\theta = \frac{\Gamma(\gamma+2)\Gamma(3/2)}{\Gamma(\gamma+5/2)}\,.
\eeq
Now, using  these expressions in the definition of $\beta(r)$ one finds
\beq
\beta(r)= \frac{r^{2} - r_a^2\gamma}{r^2+r_a^2} \,,
\eeq
which in the limit $r_a\rightarrow+\infty$ becomes
$\beta(r)=-\gamma$ while in the limit $\gamma\rightarrow 0$ one
recovers the Osipkov-Merritt anisotropy parameter given in
Eq.~\eqref{eq:OMbeta}.  This proves the expressions for $\beta(r)$
given in section~\ref{sec:OM}. Using Eqs.~(\ref{eq:sigmar}) and
(\ref{eq:sigmatheta}) in their present form one can also explicitly
evaluate Eq.~(\ref{eq:betalinear}).

For the phase space density ansatz used in
section~\ref{sec:alt-models} in general the velocity dispersions have
to be calculated numerically. Explicitly the integrals are given as
\begin{align}
  \rho \sigma^2_t = \int d^3v \, v_t^2 \, f(\eps, L) & =
4\pi  \int_0^{\sqrt{2\Psi(r)}} dv \int_0^v dv_t 
            \frac{v \, v_t^3}{\sqrt{v^2 - v_t^2}}  f(\eps, L) \\
&=4\pi \int_0^{\Psi} d\eps [2(\Psi - \eps)]^{3/2} k(\eps) \int_0^1 du 
            (1-u^2) h(\eps, L) 
\end{align}
and
\begin{align}
  \rho \sigma^2_r = \int d^3v \, v_r^2 \, f(\eps, L) & =
4\pi \int_0^{\sqrt{2\Psi(r)}} dv \int_0^v dv_t 
            \, v \, v_t \sqrt{v^2 - v_t^2}  f(\eps, L) \\
&=4\pi \int_0^{\Psi} d\eps [2(\Psi - \eps)]^{3/2} k(\eps) \int_0^1 du 
            u^2 \, h(\eps, L) 
\end{align}
with $L = \sqrt{2(\Psi - \eps)(1-u^2)} r(\Psi)$. This involves the numerical 
calculation of double integrals.


\end{document}